\documentclass[preprint,review]{elsarticle}

\journal{Journal of Systems and Software} 

\usepackage{graphicx}
\graphicspath{{./fig/}}
\DeclareGraphicsExtensions{.pdf .jpg .eps}

\usepackage[english]{babel}
\usepackage{t1enc}
\usepackage[T1]{fontenc}
\usepackage{booktabs}
\usepackage{rotating}
\usepackage{tablefootnote}
\usepackage{boldline}

\usepackage[multiple]{footmisc}

\usepackage{footnote}
\usepackage{multirow}
\usepackage{amsmath}
\usepackage[hidelinks]{hyperref}

\usepackage{listings}

\usepackage[table]{xcolor}

\usepackage{tabularx}
\newcolumntype{C}[1]{>{\centering\let\newline\\\arraybackslash\hspace{0pt}}m{#1}}
\newcolumntype{X}[1]{>{\raggedright\arraybackslash}m{#1}}

\colorlet{punct}{red!60!black}
\definecolor{background}{HTML}{EEEEEE}
\definecolor{delim}{RGB}{20,105,176}
\colorlet{numb}{magenta!60!black}

\usepackage{url}
\usepackage[backgroundcolor=blue!10,bordercolor=gray]{todonotes}

\newcommand{\gcell}{\cellcolor[rgb]{ .851.  .851.  .851}}

\let\OLDthebibliography\thebibliography
\renewcommand\thebibliography[1]{
  \OLDthebibliography{#1}
  \setlength{\parskip}{0pt}
  \setlength{\itemsep}{0pt plus 0.3ex}
}

\usepackage{float}

\begin{document}
\begin{frontmatter}

\title{An Automatically Created Novel Bug Dataset\\and its Validation in Bug Prediction}

\author[sed]{Rudolf Ferenc\corref{mycorrespondingauthor}}
\cortext[mycorrespondingauthor]{Corresponding author. Postal address: H-6720 Szeged, Dugonics tér 13, Hungary.}
\ead{ferenc@inf.u-szeged.hu}

\author[sed]{P\'eter Gyimesi}
\ead{pgyimesi@inf.u-szeged.hu}

\author[sed]{G\'abor Gyimesi}
\ead{ggyimesi@inf.u-szeged.hu}

\author[sed]{Zolt\'an T\'oth}
\ead{zizo@inf.u-szeged.hu}

\author[sed,rgai]{Tibor Gyim\'othy}
\ead{gyimothy@inf.u-szeged.hu}

\address[sed]{Department of Software Engineering, University of Szeged, Hungary}
\address[rgai]{MTA-SZTE Research Group on Artificial Intelligence, Szeged, Hungary}

\begin{abstract}

Bugs are inescapable during software development due to frequent code changes, tight deadlines, etc.; therefore, it is important to have tools to find these errors.
One way of performing bug identification is to analyze the characteristics of buggy source code elements from the past and predict the present ones based on the same characteristics, using e.g. machine learning models.
To support model building tasks, code elements and their characteristics are collected in so-called bug datasets which serve as the input for learning.

We present the \emph{BugHunter Dataset}: a novel kind of automatically constructed and freely available bug dataset containing code elements (files, classes, methods) with a wide set of code metrics and bug information.
Other available bug datasets follow the traditional approach of gathering the characteristics of all source code elements (buggy and non-buggy) at only one or more pre-selected release versions of the code.
Our approach, on the other hand, captures the buggy and the fixed states of the same source code elements from the narrowest timeframe we can identify for a bug's presence, regardless of release versions.
To show the usefulness of the new dataset, we built and evaluated bug prediction models and achieved F-measure values over 0.74.


\end{abstract}

\end{frontmatter}

\section{Introduction} \label{introduction}

The characterization of buggy source code elements is a popular research area these days.
Programmers tend to make mistakes despite the assistance provided by different integrated development environments, and errors may also occur due to frequent changes in the code and inappropriate specifications; therefore, it is important to get more and/or better tools to help the automatic detection of errors \cite{johnson2013don}.
For automatic recognition of unknown faulty code elements, it is a prerequisite to characterize the already known ones. 

During software development, programmers use a wide variety of tools, including bug tracking, task management, and version control systems.
There are numerous commercial and open-source software systems available for these purposes.
Furthermore, different web services are built to meet these needs.
The most popular ones like SourceForge, Bitbucket, and GitHub fulfill the above mentioned functionalities.
They usually provide several services, such as source code hosting and user management.
Their APIs make it possible to retrieve various kinds of data, e.g., they provide support for the examination of the behavior or the co-operation of users or even for analyzing the source code itself.
Since most of these services include bug tracking, it raises the idea to use this information in the characterization of buggy source code parts~\cite{bugcenter}.
To do so, the bug reports managed by these source code hosting providers must be connected to the appropriate source code parts~\cite{wu2011relink}.
A common practice in version control systems is to describe the changes in a comment belonging to a commit (log message) and often provide the identifier of the associated bug report which the commit is supposed to fix~\cite{githubpromises}.
This can be used to identify the faulty versions of the source code~\cite{ibugs,dallmeier2007extraction}.
Processing diff files can help us obtain the code sections affected by the bug~\cite{versioncontrol}.
We can use source code metrics~\cite{metrichistorybug}, for which we only need a tool that is able to produce them.

To build a dataset containing useful buggy code element characterization information, we chose GitHub, since it hosts several regularly maintained projects and also a well defined API that makes it possible to implement an automatic data retrieval tool.
We selected 15 Java projects as our subject systems, which differ in many ways from each other (size, domain, number of bugs reported) to cover a wide and general set of systems.

Previously published datasets follow a traditional concept to create a dataset that serves as a benchmark for testing bug prediction techniques~\cite{bugprediction,jureczko2010towards}.
These datasets include all code elements~--~both buggy and non-buggy~--~from one or more versions of the analyzed system.
In this work, we created a new approach that collects before-fix and after-fix snapshots of source code elements (and their characteristics) that were affected by bugs and does not consider those code elements that were not touched by bugs.
This kind of dataset helps to capture the changes in software product metrics when a bug is being fixed.
Thus, we can learn from the differences in source code metrics between faulty and non-faulty code elements.
As far as we know, there exists no other bug dataset yet that tries to capture this before-fix and after-fix state.

Our new dataset is called \emph{BugHunter Dataset} and it is freely available (see Section~\ref{sec:dataset}).
It can serve as a new kind of benchmark for testing different bug prediction methods since we included a wide range of source code metrics to describe the previously detected bugs in the chosen systems.
We took all reported bugs stored in the bug tracking system into consideration.
We used the usual methodology of connecting commits to bugs by analyzing the log messages and by looking for clues that would unambiguously identify the bug that was the intended target of the corresponding fixing commit(s).
Commit diffs helped us detect which source code elements were modified by a given change set, thus the code elements which had to be modified in order to fix the bug.

The first version of this work was published in our earlier conference paper~\cite{gyimesi2015characterization}.
Since then, we have extended the list of subject projects (one project was replaced because it contained template files with \textit{.java} extension which undermined the source code analysis) and we have also expanded the time interval of the analysis of the projects' history to cover three additional years (from 2014 to 2017).
Furthermore, we have refined our technique and we have incorporated method level bug information into our dataset as well.
The dataset we present in this paper has not been published before.

Here we also performed experiments to check whether our novel dataset is suitable for bug prediction purposes.
During this investigation, we collected bug characterization metrics at three source code levels (file, class, method).
After the dataset was constructed, we used different machine learning algorithms to analyze the usefulness of the dataset.

We also performed a novel kind of experiment in which we assessed whether the method level metrics are better predictors when projected to class level than the class level metrics themselves.

An important aspect to investigate is how the bug prediction models built from the novel dataset compare to the ones which used the traditional datasets as corpus.
However, this comparison is hard in its nature due to the variability in multiple factors.
One major problem is the difference in the corpus itself.
The list of the included projects vary from dataset to dataset.
In our previous work, we constructed a traditional dataset, the \emph{GitHub Bug Dataset}~\cite{toth2016public}, which consists of the same 15 projects we also included in our novel bug dataset.
This gives an opportunity to assess if there is any difference in the bug prediction capabilities of these two datasets.

To emphasize the research artifact contribution and the research questions, we highlighted them in the following box.

\bigskip
\noindent
\fbox{
  \parbox{\textwidth-16pt}{
    \textbf{Research artifact:} A freely available novel dataset containing source code metrics of buggy Java source code elements (file, class, method) before and after bug fixes were applied to them.
        
    \smallskip
    \textbf{RQ1:} Is the constructed dataset usable for bug prediction purposes?
        
    \smallskip
    \textbf{RQ2:} Are the method level metrics projected to class level better predictors than the class level metrics themselves?

    \smallskip
    \textbf{RQ3:} Is the \emph{BugHunter Dataset} more powerful and expressive than the \emph{GitHub Bug Dataset}, which is constructed with the traditional approach?
  }
}
\bigskip

The rest of the paper is structured as follows.
In Section~\ref{related} we discuss related work.
We present some statistics about GitHub and the projects that we have chosen for this work in Section~\ref{datasource}.
Next, we introduce the metrics used for bug characterization in Section~\ref{metrics}.
We describe our approach for generating and validating the dataset in detail in Section~\ref{approach}.
Then, we evaluate it by applying different machine learning algorithms in Section~\ref{evaluation}, where we also address our research questions.
In Section~\ref{threats}, we present the threats to validity.
Finally, we conclude and describe some future work directions in Section~\ref{conclusion}.

\section{Related Work} \label{related}

\subsection{Bug Localization and Source Code Management}

Many approaches have been presented dealing with bug characterization and localization~\cite{saha2013improving,wang2014version,davies2012using}.
Zhou et al. published a study describing BugLocator~\cite{bugcenter}, a tool that detects the relevant source code files that need to be changed in order to fix a bug.
BugLocator uses textual similarities (between initial bug report and the source code) to rank potentially fault-prone files.
Prior information about former bug reports is stored in a bug database.
Ranking is based on the assumption that descriptions with high similarities imply that the related files are highly similar too.
A similar ranking is done by Rebug-Detector~\cite{bugsfromsourcecode}, a tool made by Wang et al. for detecting related bugs from source code using bug information.
The tool focuses on overridden and overloaded method similarities.
In our study, we constructed a dataset that stores information about formerly buggy code elements that are now fixed, thus the same method could be applied by using source code metrics for detecting similar source code elements that are possibly fault-prone.

ReLink~\cite{wu2011relink} was developed to explore missing links between code changes committed in version control systems and the fixed bugs.
This tool could be helpful for software engineering research based on linkage data, such as software defect prediction.
ReLink mines and analyzes information like bug reporter, description, comments, and date from a bug tracking database and then tries to pair the bug with the appropriate source code files based on the set of source code information extracted from a version control system.
Most of the studies dealing with this kind of linkage data use the SZZ algorithm, which has been improved over the years \cite{kim2006automatic,williams2008szz}.
This approach uses file level textual features to extract extra information between bugs and the source code itself.
We characterized the set of bugs in a given system at file, class, and method levels by assigning different source code metrics to the source code elements (methods, classes, and files) affected by bugs.
Other approaches try to link information retrieved from only version control and bug tracking systems \cite{fischer2003populating,mockus2000identifying}.

Kalliamvakou et al. mined GitHub repositories to investigate their characteristics and their qualities~\cite{githubpromises}.
They presented a detailed study discussing different project characteristics, such as (in)activity, while also involving further research questions -- e.g., whether a project is standalone or part of a more massive system.
Results showed that the extracted data can serve as a good input for various investigations, however one must use them with caution and always verify the usefulness and reliability of the data.
It is a good practice to choose projects with many developers and commits, but it should always be kept in mind that the most important point is to choose projects that fit well for one's own purpose.
In our case, we have tried to create a dataset that is large, reliable (through some manual validation) and general enough for testing different bug prediction techniques \cite{catal2009systematic,porter1990empirically,ostrand2005predicting,ma2006statistical,zhou2006empirical}, while still being created in an automatic way.

Mining software repositories can be a harsh task when an automatic mechanism is used to construct a large set of data based on the information gathered from a distributed software repository.
As we used GitHub to address our research questions, we paid extra attention to prevent and avoid pitfalls.
Bird et al.~\cite{mininggit} presented a study on distributed version control systems -- focusing mainly on Git -- that examined their usage and the available set of data (such as whether the commits are removable, modifiable, movable).
The main purpose of the paper was to draw attention to pitfalls and help researchers to avoid them during the processing and analysis of a mined information set.

Many research papers showed that using a bug tracking system improves the quality of the software system under development.
Bangcharoensap et al. introduced a method to locate the buggy files in a software system very quickly using the bug reports managed by the bug tracking system~\cite{eclipsebugs}.
The presented method contains three different approaches to rank the fault-prone files, namely: (a) Text mining, which ranks files based on the textual similarity between a bug report and the source code itself, (b) Code mining, which ranks files based on prediction of the potential buggy module using source code product metrics, and (c) Change history, which ranks files based on prediction of the fault-prone module using change process metrics.
They used the gathered project data collected on the Eclipse platform to investigate the efficiency of the proposed approaches and showed that they are indeed suitable to locate buggy files.
Furthermore, bug reports with a short description and many specific words greatly increase the effectiveness of finding the ``weak points'' of the system.

Similarly to our study, Ostrand et al. investigated fault prediction by using source code metrics.
However, only file level was considered as the finest granularity unit~\cite{ostrand2007automating}, while we have built a toolchain to also support class and method levels.

In addition to the above presented methods, a significant change in source code metrics can also indicate that the relevant source code files contain potential bugs \cite{gyimothy2005empirical}.
Couto et al. presented a paper that shows the possible relationship between changed source code metrics (used as predictors) and bugs as well~\cite{metrichistorybug}.
They described an experiment to discover more robust evidences towards causality between software metrics and the occurrence of bugs.
Although our method does not include this specific information, we still aim to show that considering methods as basic elements and including them in a dataset is also a way for building a working corpus for bug prediction techniques.

\subsection{Public Datasets}

The previously mentioned approaches use self-made datasets for their own purposes, as illustrated in the work of Kalliamvakou et al. too~\cite{githubpromises}.
Bug prediction techniques and approaches can be presented and compared in different ways; however, there are some basic points that can serve as common components~\cite{li2018progress}.
One common element can be a dataset used for the evaluation of the various approaches.
\emph{PROMISE}
\cite{shirabad2005promise} is a repository of datasets out of which several ones contain bugs gathered from open-source and also from closed-source industrial software systems.
Amongst others it includes the NASA MDP dataset, which was used in many research studies and also criticized for containing erroneous data~\cite{shepperd2013data,petric2016jinx}.
The PROMISE repository also contains an extensively referenced dataset created by Jureczko~\cite{jureczko2010towards}, which provides object-oriented metrics as well as bug information for the source code elements (classes).
This latter one includes open-source projects such as Apache Ant, Apache Camel, JEdit, Apache Lucene, forming a dataset containing 48 releases of 15 projects.
The main purpose of these datasets is to support prediction methods and summarize bugs and their characterizations extracted from various projects.
Many research papers used datasets from the PROMISE repository as an input for their investigations.

A similar dataset for bug prediction came to be commonly known as the \emph{Bug prediction dataset\footnote{\url{http://bug.inf.usi.ch/}}}~\cite{bugprediction}.
The reason for creating this dataset was mainly inspired by the idea of measuring the performance of the different prediction models and also comparing them to each other.
This dataset handles the bugs and the relevant source code parts at class level, i.e., the bugs are assigned to classes.
As we have already mentioned, we do not only focus on file and class levels, but on method-level elements as well.

Zimmermann et al.~\cite{zimmermann2007predicting} used Eclipse as the input for a study dealing with defect prediction.
They investigated whether the complexity metrics have the power to detect fault prone points in the system at package and file level.
During the study, they constructed a public dataset\footnote{\url{https://www.st.cs.uni-saarland.de/softevo/bug-data/eclipse/}} that is still available.
It contains different source code metrics and a subset of the files/packages is marked as ``buggy'' if it contained any bugs in the interval between two releases.

A recent study showed that code smells also play a significant role in bug prediction \cite{hall2014some} but the constructed dataset is not public.
In our dataset, we also include code smell metrics to enhance its usefulness.

\emph{iBUGS}~\cite{dallmeier2007extraction} provides a complex environment for testing different automatic defect localization methods.   
Information describing the bugs comes from both version control systems and from bug tracking systems.
iBUGS used the following three open-source projects to extract the bugs from (the numbers of extracted bugs are in parentheses):
	AspectJ -- an extension for the Java programming language to support aspect oriented programming (223);
	Rhino -- a JavaSript interpreter written in Java (32); and
	Joda-Time -- a quality replacement (extension) for the Java date and time classes (8).
The authors attempted to generate the iBUGS dataset in an automatic way and they compared the generated set to the manually validated set of bugs~\cite{ibugs}.
iBUGS is a framework aimed more towards bug localization and not a standalone dataset containing source code elements and their characterizations (i.e., metrics).

The \emph{Bugcatchers}~\cite{hall2014some} dataset is created by Hall et al. which is not only a bug dataset, but also contains bad smells detected in the subject systems.
The selected three systems are Eclipse JDT Core, ArgoUML, and Apache Commons.
The dataset is built and evaluated at file level.

The \emph{ELFF} dataset~\cite{Shippey2016Esem} is a recent dataset proposed by Shippey et al.
They experienced that only a few method level datasets exist, thus they created a dataset whose entries are methods.
Additionally, they also made class level datasets publicly available.
They used Boa~\cite{dyer2013boa} to mine SourceForge repositories and collect as many candidates as they can, selecting 23 projects out of 50,000 that fulfilled their criteria (number of closed bugs, bugs are referenced from commits, etc.).
They only kept projects with SVN version control systems which narrows down their candidate set.
They used the classic and well-defined SZZ algorithm~\cite{sliwerski2005changes} to find linkage between bugs and the corresponding source code elements.

The \emph{Had-oops!} dataset~\cite{Harman:2014ssbse} is constructed by a new approach presented by Harman et al.
They analyzed 8 consecutive Hadoop versions and investigated the impact of chronology on fault prediction performance.
They used Support Vector Machines (SVMs) with the Genetic Algorithm (for configuration) to build prediction models at class level.
For a given version, they constructed a prediction model from all the previous versions and a model from only the current version and compared which one performed better.
Results are not straightforward since they found early versions preferable in many cases as opposed to models built on recent versions.
Moreover, using all versions is not always better than using only the current version to build a model from.

The \emph{Mutation-aware fault prediction dataset} is a result of an experiment carried out by Bowes et al. on using mutation metrics as independent variables for fault prediction~\cite{Bowes:2016ISSTA}.
They used 3 software systems from which 2 projects (Eclipse and Apache) were open-source and one was closed.
They used the popular PITest (or simply, PIT~\cite{pitest}) to obtain the set of mutation metrics that were included in the final dataset.
Besides the mutation metrics, some static source code metrics (calculated by JHawk~\cite{jhawk}) were also included in the dataset for comparison purposes.
This dataset is also built at class level.

The \emph{GitHub Bug Dataset} is a recent dataset that includes class and file level static source code metrics~\cite{toth2016public} for 15 Java systems gathered from GitHub.
Besides size, documentation, object-oriented, and complexity metrics, the dataset also contains code duplication and coding rule violation metrics.
This dataset is our previous work that was still constructed in the ``traditional'' way.
In Table~\ref{tab:datasetcharacteristics}, we compare the main characteristics of the mentioned datasets.

Our goal was to pick the strong aspects of all the previous datasets and put them together, as its positive effects are described by Li et al.~\cite{li2019multiple}.
Although the discussed works successfully made use of their datasets, an extended dataset can serve as a good basis for further investigations.
Our dataset includes various projects from GitHub and includes numerous static source code metrics and stores a large number of entries in fine granularity (file, class, and method level as well).
Furthermore, we also experimented with chronology, although in a different way compared to Harman et al~\cite{Harman:2014ssbse}.
The differences between the traditional datasets and the proposed novel dataset are summarized in Table~\ref{tab:datasetcomparison}.
See Section~\ref{sec:processingrawdata} for details about the process of selecting the bug related data for the novel dataset.
The detailed comparison can be found in Section~\ref{sec:rq3}.

\begin{table}[H]
\caption{Comparison of the two types of datasets}
\fontsize{7pt}{8pt}\selectfont
\centering
\setlength{\tabcolsep}{3pt}
\begin{tabular}{|l|C{3.8cm}|C{3.8cm}|}
\hline
\textbf{Feature} & \textbf{Traditional} & \textbf{Novel} \\
\hline
\textit{Included time interval} & Usually 6 months & Entire project history \\
\hline
\textit{Included source code elements} & All the elements from a single version & Only the modified elements right before and after bug-fixes \\
\hline
\textit{Assumptions} & Source code elements that are not included in any bug-fix are non-faulty & No assumptions needed \\
\hline
\textit{Uncertainty} & The source code elements are faulty in the latest release version before the bug-fix and non-faulty after the fix & The source code elements are faulty right before the bug-fix and fixed afterwards \\
\hline
\end{tabular}
\label{tab:datasetcomparison}
\end{table}

\begin{table}[H]
\caption{Comparison of the datasets}
\fontsize{7pt}{8pt}\selectfont
\centering
\setlength{\tabcolsep}{3pt}
\begin{tabular}{|X{3,5cm}|l|X{3,8cm}|l|l|}
\hline
\textbf{Project}                                 & \textbf{Level of bugs} & \textbf{Bug characteristics} & \textbf{\# of projects} \\ \hline
\emph{NASA MDP Dataset}                        & class & static source code metrics &  11 \\ \hline
\emph{Jureczko Dataset}                        & class & static source code metrics &  15 \\ \hline
\emph{Bug prediction dataset}                  & class & static source code metrics, process metrics & 5 \\ \hline
\emph{Eclipse dataset}                         & file, package & complexity metrics & 1 \\ \hline
\emph{iBUGS}                                   & N/A & bug-fix size properties, AST fingerprints & 3 \\ \hline
\emph{Bugcatchers}                             & file & code smells & 3 \\ \hline
\emph{ELFF}                                    & class, method & static source code metrics & 23 \\ \hline
\emph{Had-oops!}                               & class & static source code metrics & 1 \\ \hline
\emph{Mutation-aware fault prediction dataset} & class & static source code metrics, mutation metrics & 3 \\ \hline
\emph{GitHub Bug Dataset}                      & file, class & static source code metrics, code duplication metrics, code smell metrics & 15 \\ \hline
\emph{Novel dataset}                           & file, class, method & static source code metrics, code duplication metrics, code smell metrics & 15 \\ \hline
\end{tabular}
\label{tab:datasetcharacteristics}
\end{table}

\section{Data Source} \label{datasource}

To address our research objectives, this section briefly introduces the version control system used (Git), its corresponding source code hosting service (GitHub), and their main functionalities that are closely related to the creation of linkage data between bugs and source code elements.
Afterwards, we enumerate the chosen projects and give some further insight on the reasons why we chose them as our subject systems.

\subsection{GitHub}

GitHub is one of today's most popular source code hosting services.
It is used by several major open-source teams for managing their projects like Node.js, Ruby on Rails, Spring Framework, Zend Framework, and Jenkins, among others.
GitHub offers public and private Git repositories for its users, with some collaborative services, e.g., built-in bug and issue tracking systems.

Bug reporting is supported by the fact that any GitHub user can add an issue, and collaborators can even label these issues for further categorization.
The system provides some basic labels, such as ``bug'', ``duplicate'', and ``enhancement'', but these tags can be customized if required.
In an optimal case, the collaborators review these reports and label them with the proper labels, for instance, the bug reports with the ``bug'' label.
The most important feature of bug tracking is that we can refer to an issue from the log message of a commit by using the unique identifier of the issue, thereby identifying a connection between the source code and the reported bug.
GitHub has an API\footnote{\url{https://developer.github.com/v3/}} that can be used for managing repositories from other systems, or query information about them.
This information includes events, feeds, notifications, gists, issues, commits, statistics, and user data.

\begin{table}[h]
\caption{The number of repositories created between 01-01-2013 and 07-09-2017 for the top~10 languages}
\def\arraystretch{1}
\centering
\setlength{\tabcolsep}{6pt}
\begin{tabular}{l|c}
\hline
\textbf{Language} & \textbf{Number of repositories} \\ \hline
JavaScript     & 2,019,215              \\ 
Java           & 1,465,168              \\ 
Ruby           & 1,379,225              \\ 
Python         & 1,014,760              \\ 
PHP            & 983,479                \\ 
C              & 737,314                \\ 
C++            & 619,914                \\ 
CSS            & 568,493                \\ 
C\#            & 282,092                \\ 
Shell          & 263,350                \\ \hline
\end{tabular}
\label{tab:githubarchivedata}
\end{table}

With the GitHub Archive\footnote{\url{https://www.gharchive.org/}} project that also uses this API, we can get up-to-date statistics about the public repositories.
For instance, Table~\ref{tab:githubarchivedata} presents the number of repositories created between 1 January 2013 and 7 September 2017, grouped by the main programming languages they use (only the top 10 languages are shown).
Our approach uses Java repositories (the second most used platform on GitHub) to gather a proper information base for constructing a bug dataset.

Although extracting basic information from GitHub is easy, some version control features are hard to deal with, especially during the linking process when we try to match source code elements to bugs.
For example, Git provides a powerful branching mechanism by supporting the creation, deletion, and selection of branches.
In our case, we have to handle different branches because a fixing commit most often occurs on other -- so called ``topic'' -- branches and not on the master branch.
Fortunately, the projects we analyzed often solved this problem by merging.
During the merge, isomorphic commits are generated and placed on the master branch, thus all the desired analysis can be done by taking only the master branch with a given version as input.
Another example is forking a repository, which is used world-wide.
In our experiment, we do not handle forks, since it would have encumbered the above mentioned linking process and we would not gain significant additional information since bugs are often corrected in the original repository.
These details can be viewed as our underlying assumptions regarding the usage of GitHub.

\subsection{The Chosen Projects}\label{sec:chosenprojects}

We considered several criteria when searching for appropriate projects on GitHub.
First of all, we searched for projects written in Java, especially larger ones, because those are more suitable for this kind of analysis.
It was also important to have an adequate number of issues labeled as bugs, and the number of references from the log messages to certain commits is also a crucial factor (this is how we can link source code elements to bugs).
Additionally, we preferred projects that are still actively maintained.
Logged-in users can give a star for any repository and bookmark selected ones to follow.
The number of stars and watches applied to repositories forms a ranking between them, which we will refer to as ``popularity'' in the following.
We performed our search for candidate projects mainly based on popularity and activity.
We also found many projects during the search that would have fulfilled most aspects, had the developers not used an external bug tracker system -- something we could not support yet.

In the end, we selected the 15 projects listed in Table~\ref{tab:projects} based on the previously mentioned criteria.
As the descriptions show, these projects cover different domains; a good practice when the goal is creating a general dataset.
The table contains the following additional data about the projects:
\begin{description}
  \item[Stars] the number of stars a project received on GitHub
  \item[Forks] the number of forks of a project on GitHub
  \item[kLOC] the thousand lines of code a project had at September, 2017
\end{description}
Recently, the repository of the Ceylon project was moved to a new location and the old repository is not available anymore.
Due to this reason we could not obtain the total number of stars and the total number of forks of this repository, resulting the low values in the table.

\begin{table}[htb!]
\caption{The selected projects and their descriptions}
\scriptsize
\def\arraystretch{1}
\centering
\setlength{\tabcolsep}{6pt}
\begin{tabular}{|C{1,5cm}C{1,5cm}C{1,5cm}|p{7cm}|}
\hline
\multicolumn{3}{|c|}{\textbf{Project Name}} & \multirow{2}{*}{\textbf{Description}} \\
\textbf{Stars} & \textbf{Forks} & \textbf{kLOC} &  \\ \hline

\multicolumn{3}{|c|}{Android Universal Image Loader\tablefootnote{\url{https://github.com/nostra13/Android-Universal-Image-Loader}}} & \multirow{2}{=}{An Android library that assists the loading of images.}\\ 
 16,521 & 6,357 & 13 & \\ \hline

\multicolumn{3}{|c|}{ANTLR v4\tablefootnote{\url{https://github.com/antlr/antlr4}}} & \multirow{3}{=}{A popular software in the field of language processing. It is a powerful parser generator for reading, processing, executing, or translating structured text or binary files.}\\ 
\multirow{2}{*}{6,030} & \multirow{2}{*}{1,559} & \multirow{2}{*}{68} & \\
 &  &  & \\ \hline

\multicolumn{3}{|c|}{Elasticsearch\tablefootnote{\url{https://github.com/elastic/elasticsearch}}} & \multirow{2}{=}{A popular RESTful search engine.}\\ 
 42,685 & 14,303 & 995 & \\ \hline

\multicolumn{3}{|c|}{jUnit\tablefootnote{\url{https://github.com/junit-team/junit4}}} & \multirow{2}{=}{A Java framework for writing unit tests.}\\ 
 7,536 & 2,826 & 43 & \\ \hline

\multicolumn{3}{|c|}{MapDB\tablefootnote{\url{https://github.com/jankotek/MapDB}}} & \multirow{2}{=}{A versatile, fast and easy to use database engine in Java.}\\ 
 3,700 & 745 & 68 & \\ \hline

\multicolumn{3}{|c|}{mcMMO\tablefootnote{\url{https://github.com/mcMMO-Dev/mcMMO}}} & \multirow{2}{=}{An RPG game based on Minecraft. }\\ 
 511 & 448 & 42 & \\ \hline

\multicolumn{3}{|c|}{Mission Control Technologies\tablefootnote{\url{https://github.com/nasa/mct}}} & \multirow{4}{=}{Originally developed by NASA for the space flight operations. It is a real-time monitoring and visualization platform that can be used for monitoring any other data as well.}\\ 
 \multirow{3}{*}{818} & \multirow{3}{*}{280} & \multirow{3}{*}{204} & \\ 
  &  &  & \\ 
  &  &  & \\ \hline

\multicolumn{3}{|c|}{Neo4j\tablefootnote{\url{https://github.com/neo4j/neo4j}}} & \multirow{2}{=}{The world's leading graph database with high performance.}\\ 
 6,643 & 1,636 & 1,027 & \\ \hline

\multicolumn{3}{|c|}{Netty\tablefootnote{\url{https://github.com/netty/netty}}} & \multirow{2}{=}{An asynchronous event-driven networking framework.}\\ 
 20,006 & 9,128 & 380 & \\ \hline

\multicolumn{3}{|c|}{OrientDB\tablefootnote{\url{https://github.com/orientechnologies/orientdb}}} & \multirow{2}{=}{A popular document-based NoSQL graph database. Mainly famous for its speed and scalability.}\\ 
 3,919 & 792 & 621 & \\ \hline

\multicolumn{3}{|c|}{Oryx 2\tablefootnote{\url{https://github.com/OryxProject/oryx}}} & \multirow{2}{=}{An open-source software with machine learning algorithms that allows the processing of huge data sets.}\\ 
 1,633 & 388 & 34 & \\ \hline

\multicolumn{3}{|c|}{Titan\tablefootnote{\url{https://github.com/thinkaurelius/titan}}} & \multirow{2}{=}{A high-performance, highly scalable graph database.}\\ 
 4,931 & 1,015 & 108 & \\ \hline

\multicolumn{3}{|c|}{Eclipse plugin for Ceylon\tablefootnote{\url{https://github.com/eclipse/ceylon-ide-eclipse}}} & \multirow{2}{=}{An Eclipse plugin which provides a Ceylon IDE.}\\ 
 56 & 30 & 181 & \\ \hline

\multicolumn{3}{|c|}{Hazelcast\tablefootnote{\url{https://github.com/hazelcast/hazelcast}}} & \multirow{2}{=}{A platform for distributed data processing.}\\ 
 3,211 & 1,169 & 949 & \\ \hline

\multicolumn{3}{|c|}{Broadleaf Commerce\tablefootnote{\url{https://github.com/BroadleafCommerce/BroadleafCommerce}}} & \multirow{2}{=}{A framework for building e-commerce websites.}\\ 
 1,266 & 1,020 & 322 & \\ \hline

\end{tabular}
\label{tab:projects}
\end{table}

\begin{table}[htb!]
\caption{Statistics about the selected projects}
\scriptsize
\def\arraystretch{1.5}
\centering
\setlength{\tabcolsep}{4pt}
\begin{tabular}{|l|rrrrrr|}
\hline
 & \textbf{TNC} & \textbf{NCRB} & \textbf{NBR} & \textbf{NOBR} & \textbf{NCBR} & \textbf{ANCBR} \\ \hline
\textbf{Android Universal I. L.} & 1,025  & 52    & 90    & 15  & 75    & 0.69 \\
\textbf{ANTLR v4}                & 6,526  & 162   & 179   & 23  & 156   & 1.04   \\
\textbf{Elasticsearch}           & 28,815 & 2,807 & 4,494 & 207 & 4,287 & 0.65 \\
\textbf{jUnit}                   & 2,192  & 72    & 90    & 6   & 84    & 0.86   \\
\textbf{MapDB}                   & 2,062  & 167   & 244   & 16  & 228   & 0.73   \\
\textbf{mcMMO}                   & 4,765  & 268   & 664   & 8   & 656   & 0.41   \\
\textbf{Mission Control T.}      & 977    & 15    & 46    & 9   & 37    & 0.40 \\
\textbf{Neo4j}                   & 49,979 & 781   & 1,268 & 116 & 1,152 & 0.68 \\
\textbf{Netty}                   & 8,443  & 956   & 2,240 & 33  & 2,207 & 0.43 \\
\textbf{OrientDB}                & 15,969 & 722   & 1,522 & 250 & 1,272 & 0.57 \\
\textbf{Oryx}                    & 1,054  & 69    & 67    & 2   & 65    & 1.06   \\
\textbf{Titan}                   & 4,434  & 93    & 135   & 8   & 127   & 0.73 \\
\textbf{Eclipse p. for Ceylon}   & 7,984  & 316   & 923   & 82  & 841   & 0.38 \\
\textbf{Hazelcast}               & 24,380 & 3,030 & 3,882 & 120 & 3,762 & 0.81 \\
\textbf{Broadleaf Commerce}      & 14,920 & 1,051 & 703   & 28  & 675   & 1.56 \\ \hline
\end{tabular}
\label{tab:projectstat}
\end{table}

Besides knowing each project's domain, further descriptors can help us get a more precise understanding.
Table~\ref{tab:projectstat} provides a more accurate picture of the projects by showing different characteristics (related to the repositories) for each project.
This table sums up the occurrences of various bug reports and commits of the projects present at September, 2017.
Considering the total number of commits (TNC) is a good starting point to show the scale and activity of the projects.
The number of commits referencing a (closed) bug (NCRB) shows how many commits out of TNC referenced a bug by using the pattern '\textit{\#x}' in their commit log messages, where \textit{x} is a number that uniquely identifies the proper issue that is labeled as a bug~\cite{mockus2000identifying}.
NCBR (Number of Closed Bug Reports) is also important, since we only consider closed bug reports and the corresponding commits in this context.
The abbreviations we used stand for the following:

\begin{description}
  \item[TNC] Total Number of Commits
  \item[NCRB] Number of Commits Referencing a Bug
  \item[NBR] Number of Bug Reports
	\item[NOBR] Number of Open Bug Reports
	\item[NCBR] Number of Closed Bug Reports
	\item[ANCBR] Average Number of Commits per closed Bug Reports (NCRB/NCBR)
\end{description}

\begin{figure}
\centering
\includegraphics[width=\textwidth]{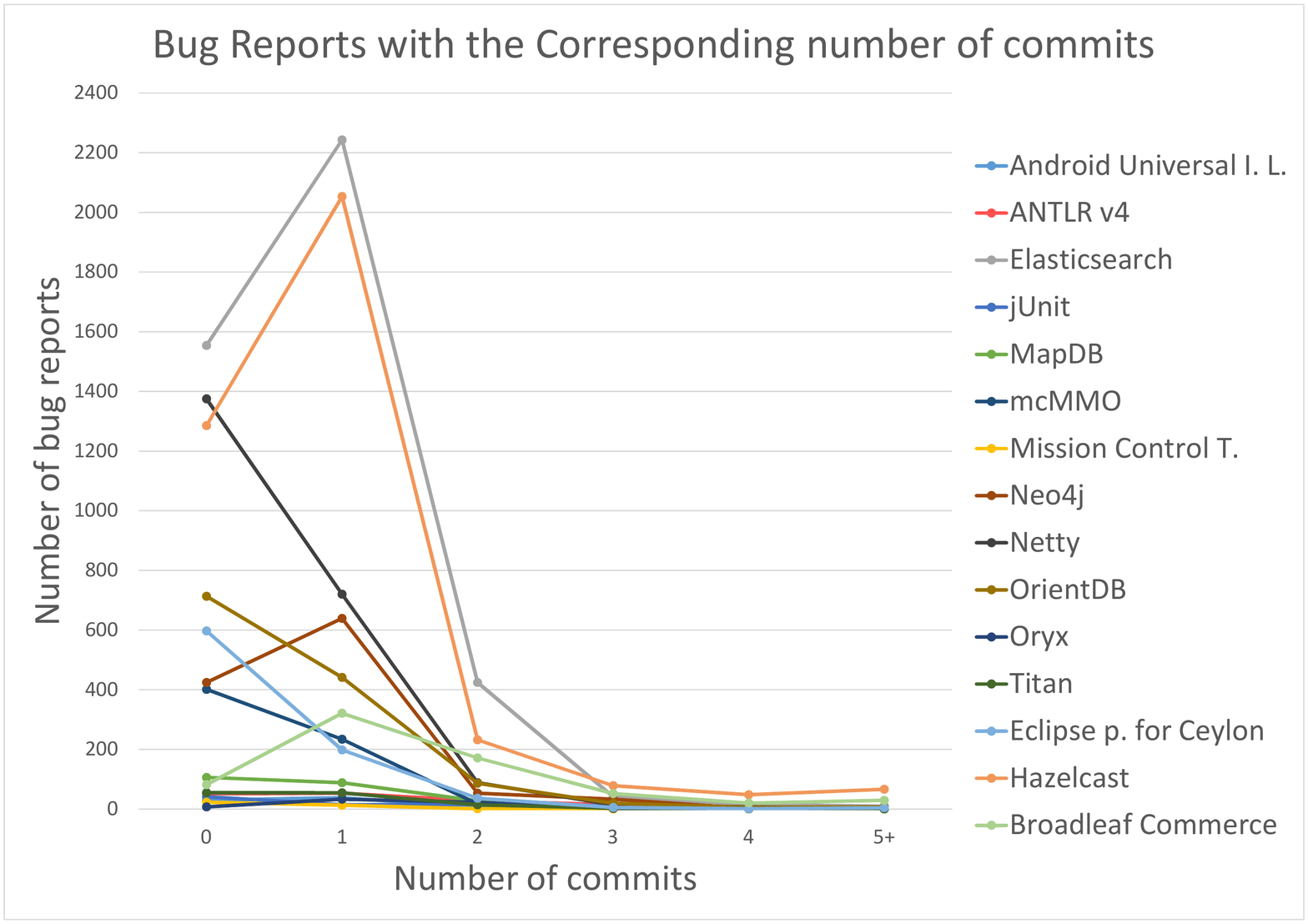}
\caption{The number of bug reports with the corresponding number of commits} 
\label{fig:issueswithncommits}
\end{figure}

It is apparent that the projects are quite different according to the number of bug reports and the lines of code they have.
NCRB is always lower than NCBR except in three cases (ANTLR v4, Oryx, Broadleaf Commerce) which means that not all bug reports have at least one referencing commit to fix the bug.
This is possible since closing a bug is viable not only from a commit but directly from GitHub's Web user interface without committing anything.

Figure \ref{fig:issueswithncommits} depicts the number of commits for each closed bug report.
One insight here is that the rate of closed bug reports is high where not even a single commit is present to fix the bug.
There are several possible causes for this, for example, the bug report is not referred from the commit's log message, or the error has already been fixed.

\begin{figure}
\centering
\includegraphics[width=\textwidth]{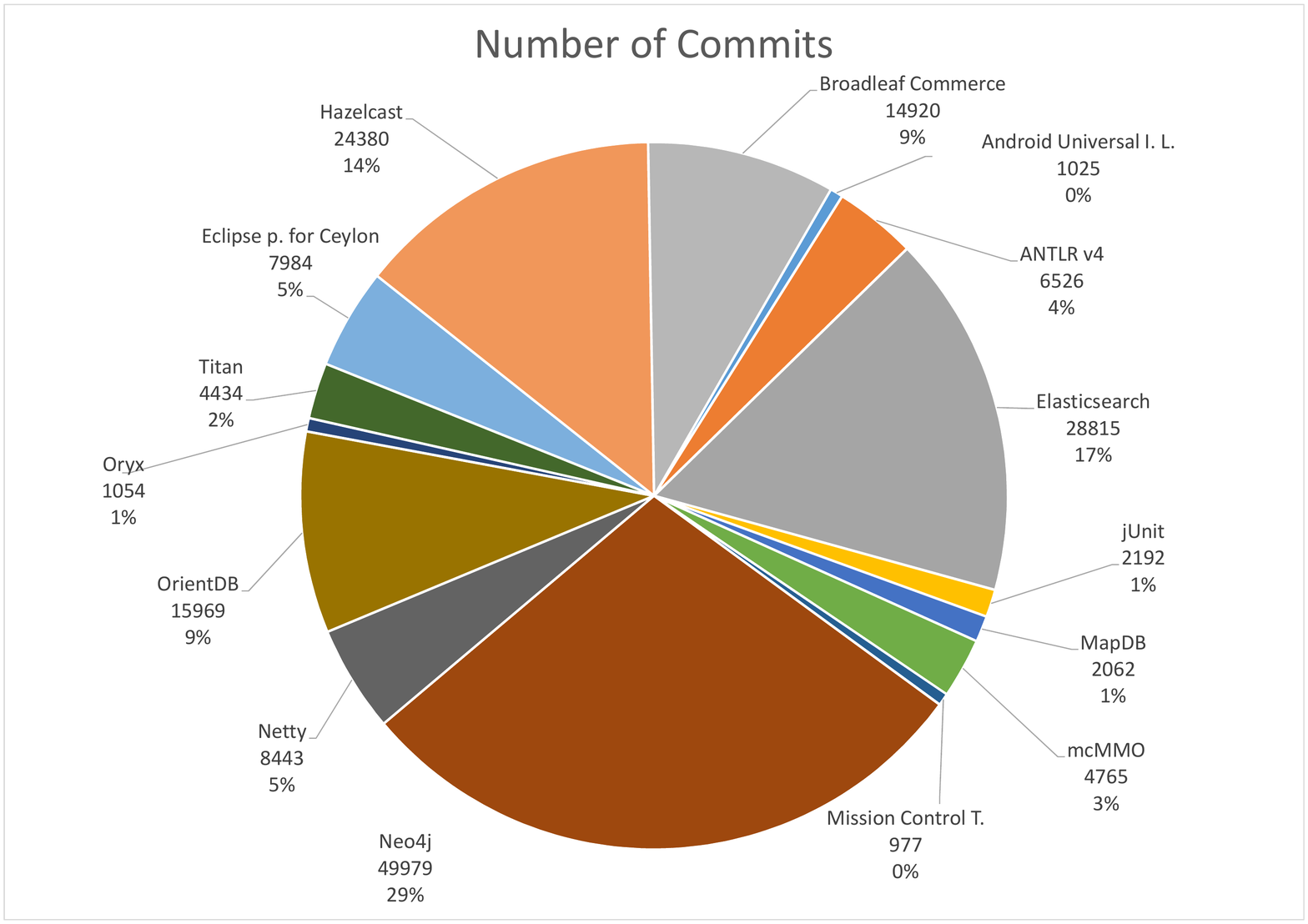}
\caption{The number of commits per projects} 
\label{fig:numberofcommits}
\end{figure}

Figure~\ref{fig:numberofcommits} shows the ratio of the number of commits per projects, illustrating the activity and the size of the projects.
Neo4j is dominant if we consider only the number of commits, however bug report related activities are slight.
The presented figures show the variability of the selected software systems, which ensures the construction of a heterogeneous dataset.

\section{Metrics}\label{metrics}

A software metric is a quantified measure of a property of a software project.
By using a set of different metrics, we can measure the properties of a project objectively from various points of view.
Metrics can be obtained from the source code, from the project management system, or even from the execution traces of the source code.
We can deduce higher-level software characteristics from lower level ones~\cite{qualitymodel}, such as the maintainability of the source code or the distribution of defects, but they can be also used to build a cost estimation model, apply performance optimization, or to improve activities supporting software quality~\cite{Boehm2000,ban2018prediction,ban2014recognizing}.
In this work, we used static source code metrics (also known as software product metrics).

The area of object-oriented source code metrics has been researched for many years~\cite{chidamber1994metrics,basili1996validation,bruntink2004predicting}, thus no wonder that several tools exist for measuring them.
These tools are suitable for detailed examination of systems written in various programming languages.
The source code metrics provide information about the size, inheritance, coupling, cohesion, or complexity of the code.
We used the \emph{OpenStaticAnalyzer}\footnote{\url{https://github.com/sed-inf-u-szeged/OpenStaticAnalyzer}} tool to obtain various software product metrics for the selected systems.
The full list of the object-oriented metrics we used is shown in Table~\ref{tab:metrics}.
The last three columns of the table indicate the kind of elements the given metric is calculated for, namely method, class, and file.
The presence of `X' indicates that the metric is calculated for the given source code level.
Most of the blanks in the table come from the fact that the metric is defined only for a given level.
For instance, \textit{Weighted Methods per Class} cannot be interpreted for methods and files.
Other blanks come from the limitations of the used static source code analyzer (i.e. OpenStaticAnalyzer).

\begin{table}[H]
\caption{Source code metrics used for characterization}
\scriptsize
\def\arraystretch{0.65}
\centering
\setlength{\tabcolsep}{6pt}
\begin{tabular}{|l|l|c|c|c|}
\hline
\textbf{Abbreviation} & \textbf{Full name}   & \textbf{Method} &  \textbf{Class} & \textbf{File} \\ \hline
CLOC & Comment Lines of Code & X & X & X\\
LOC & Lines of Code & X & X & X\\
LLOC & Logical Lines of Code & X & X & X\\
\hline
NL & Nesting Level & X & X & \\
NLE & Nesting Level Else-If & X & X & \\
NII & Number of Incoming Invocations & X & X & \\
NOI & Number of Outgoing Invocations & X & X & \\
CD & Comment Density & X & X & \\
DLOC & Documentation Lines of Code & X & X & \\
TCD & Total Comment Density & X & X & \\
TCLOC & Total Comment Lines of Code & X & X & \\
NOS & Number of Statements & X & X & \\
TLOC & Total Lines of Code & X & X & \\
TLLOC & Total Logical Lines of Code & X & X & \\
TNOS & Total Number of Statements & X & X & \\
\hline
McCC & McCabe's Cyclomatic Complexity & X &  & X\\
\hline
PDA & Public Documented API &  & X & X\\
PUA & Public Undocumented API &  & X & X\\
\hline
HCPL & Halstead Calculated Program Length & X &  & \\
HDIF & Halstead Difficulty & X &  & \\
HEFF & Halstead Effort & X &  & \\
HNDB & Halstead Number of Delivered Bugs & X &  & \\
HPL & Halstead Program Length & X &  & \\
HPV & Halstead Program Vocabulary & X &  & \\
HTRP & Halstead Time Required to Program & X &  & \\
HVOL & Halstead Volume & X &  & \\
MIMS & Maintainability Index (Microsoft version) & X &  & \\
MI & Maintainability Index (Original version) & X &  & \\
MISEI & Maintainability Index (SEI version) & X &  & \\
MISM & Maintainability Index (SourceMeter version) & X &  & \\
NUMPAR & Number of Parameters & X &  & \\
\hline
LCOM5 & Lack of Cohesion in Methods 5 &  & X & \\
WMC & Weighted Methods per Class &  & X & \\
CBO & Coupling Between Object classes &  & X & \\
CBOI & Coupling Between Object classes Inverse &  & X & \\
RFC & Response set For Class &  & X & \\
AD & API Documentation &  & X & \\
DIT & Depth of Inheritance Tree &  & X & \\
NOA & Number of Ancestors &  & X & \\
NOC & Number of Children &  & X & \\
NOD & Number of Descendants &  & X & \\
NOP & Number of Parents &  & X & \\
NA & Number of Attributes &  & X & \\
NG & Number of Getters &  & X & \\
NLA & Number of Local Attributes &  & X & \\
NLG & Number of Local Getters &  & X & \\
NLM & Number of Local Methods &  & X & \\
NLPA & Number of Local Public Attributes &  & X & \\
NLPM & Number of Local Public Methods &  & X & \\
NLS & Number of Local Setters &  & X & \\
NM & Number of Methods &  & X & \\
NPA & Number of Public Attributes &  & X & \\
NPM & Number of Public Methods &  & X & \\
NS & Number of Setters &  & X & \\
TNA & Total Number of Attributes &  & X & \\
TNG & Total Number of Getters &  & X & \\
TNLA & Total Number of Local Attributes &  & X & \\
TNLG & Total Number of Local Getters &  & X & \\
TNLM & Total Number of Local Methods &  & X & \\
TNLPA & Total Number of Local Public Attributes &  & X & \\
TNLPM & Total Number of Local Public Methods &  & X & \\
TNLS & Total Number of Local Setters &  & X & \\
TNM & Total Number of Methods &  & X & \\
TNPA & Total Number of Public Attributes &  & X & \\
TNPM & Total Number of Public Methods &  & X & \\
TNS & Total Number of Setters &  & X & \\
\hline
\end{tabular}
\label{tab:metrics}
\end{table}

One special metric category is provided by source code duplication detection~\cite{roy2009comparison}.
OpenStaticAnalyzer is able to detect Type-1 (exact copy of code, not considering white spaces and comments) and Type-2 clones (syntactically identical copy of code where variable, function or type identifiers can be different; also not considering white spaces and comments) in software systems~\cite{bellon2007comparison} and also supports clone management tasks, such as:
\begin{itemize}
	\item Clone tracking: clones are tracked during the source code analysis of consecutive revisions of the analyzed software system.
	\item Calculating clone metrics: a wide set of clone related metrics is calculated to describe the properties of a clone in the system (for example, risk of a clone or the effort needed to eliminate the clone from the system).
\end{itemize}
Basic clone related metrics that are calculated for methods and classes are presented in Table~\ref{tab:clone_metrics}.

\begin{table}[H]
\caption{Clone metrics used for characterization}
\scriptsize
\def\arraystretch{1.1}
\centering
\setlength{\tabcolsep}{6pt}
\begin{tabular}{|l|l|}
\hline
\textbf{Abbreviation} & \textbf{Full name}   \\ \hline
CC     &    Clone Coverage                    \\
CCL    &    Clone Classes                     \\
CCO    &    Clone Complexity                  \\
CI     &    Clone Instances                   \\
CLC    &    Clone Line Coverage               \\
CLLC   &    Clone Lines of Code               \\
LDC    &    Lines of Duplicated Code          \\
LLDC   &    Logical Lines of Duplicated Code  \\  \hline
\end{tabular}
\label{tab:clone_metrics}
\end{table}

OpenStaticAnalyzer also provides a coding rule violation detection module.
The presence of rule violations in a source code element can cause errors~\cite{boogerd2008assessing} in a later phase (can easily be a ticking bomb); thus the number of different rule violations located in the source code element can serve as good predictors and the dataset encapsulates this information too.

\section{Dataset Creation} \label{approach}

\begin{figure}[htb!]
\centering
\includegraphics[width=\textwidth]{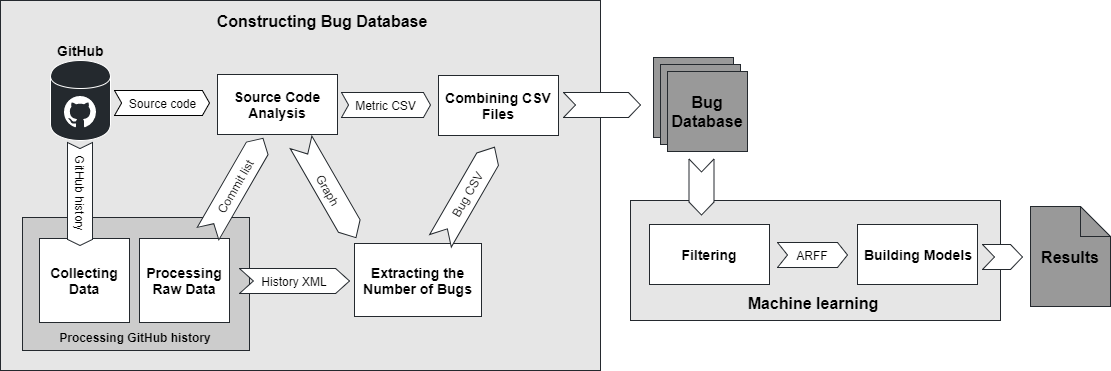}
\caption{The components of the process} 
\label{fig:toolchain}
\end{figure}

In this section, we introduce the methodology we used to create the dataset.
We carried out the data processing in multiple steps using the toolchain shown in Figure~\ref{fig:toolchain}.
Each of these steps -- and their corresponding components -- are detailed in their dedicated sections below.

\subsection{Collecting Data}

First, we save data about the selected projects via the GitHub API.
This is necessary, because while the data is continuously changing on GitHub due to the activities in the projects, we need a consistent data source for the analysis.
The data we save includes the list of users assigned to the repository (Contributors), the open and closed bug reports (Issues), and all of the commits.
For open issues, we stored only the date of their creation.
For closed issues, we stored the creation date, closing date, and the hash of the fixing commits with their commit dates.
Additionally, we focused exclusively on bug related issues, so closed bugs that were not referenced from any commit were not stored.
This filtering is based on the issue labels provided by GitHub and the set of labels we manually selected for each project.
The data we stored about the commits includes the identifier of the contributor, the parent(s) of the commit, and the affected files with their corresponding changes.
All this raw information is stored in an XML format, ready for further processing.

\subsection{Processing Raw Data} \label{sec:processingrawdata}

\begin{figure}
\includegraphics[width=\textwidth]{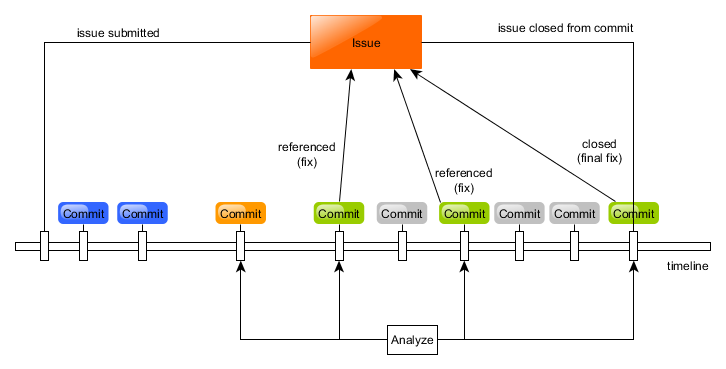}
\caption{The relationship between the bug reports and commits} 
\label{fig:issue}
\end{figure}

While the data saved from GitHub includes all commits, we only need the ones that relate to the bug reports.
These commits are then divided into different subsets, as depicted in Figure~\ref{fig:issue}.
Green nodes are directly referencing the bug report (fixing intention).
Gray nodes are commits applied between the first fix and the last fix but not referencing the bug id in their commit log messages.
One extra commit taken into consideration is the one right before the first fix (colored with orange).
This commit holds the state when the source code is buggy (not fixed yet), thus a snapshot (source code analysis) will be performed at that point too.
Although the orange node represents the latest state where the bug is not fixed yet, the blue nodes also contain the bug so we mark the source code elements as buggy in these versions too.
These blue markings are important for distinguishing commits that are involved in multiple bugs at the same time.

We have to perform code analysis on the orange and green commits to construct dataset entries.
Two entries are created for every source code element they contain: one with the state (metrics) right before the fix was applied, and one with the state when the bug was fixed.
At green commits except the last one, we do not need to perform a full code analysis, since at those points we are only interested in extracting the affected source code elements.
Amongst the selected commits, some further ones can occur that need to be removed because they are no longer available through Git (deleted, merged).
Moreover, we do not only search for links from the direction of commits but also from the direction of issues (bug reports).
When considering a bug report, we can find a commit id showing that the bug was closed in that specific commit.
At this point, the full list is constructed as a text file, which has all the commit ids (hash) for a selected project to undergo static analysis.

\subsection{Source Code Analysis} \label{sec:sourcecodeanalysis}

After gathering the appropriate versions of the source code for a given project, feature extraction can begin.
This component wraps the results of the OpenStaticAnalyzer tool that computes the source code metrics and determines the positions of the source code elements.
Results are generated in a graph format, which contains the files, classes, and methods with the computed data that includes different software product metrics (described in Section~\ref{metrics}).
At this point we have all the raw data desired, including the source code elements located in the project and all the bug related information.

\subsection{Extracting the Number of Bugs}

The next step is to link the two data sets -- the results of the code analysis and the data gathered from GitHub -- and extract the characteristics of the bugs.
Here, we determine the source code elements affected by the commits and the number of bugs in each commit for file, class, and method levels.

To determine the affected source code parts, an approach similar to the SZZ algorithm~\cite{williams2008szz} is used.
However, we do not want to detect the fix inducing commits, only the mapping between the fixing code snippets and source code elements.
For this purpose, we used the diff files -- from the GitHub data we saved -- that contain the differences between two source code versions in a unified diff format.
An example unified diff file snippet is shown below.

\begin{verbatim}
	--- /path/to/original	''timestamp''
	+++ /path/to/new	''timestamp''
	@@ -1,4 +1,4 @@
	+Added line
	-Deleted line
	 This part of the
	 document has stayed the
	 same
\end{verbatim}

Each diff contains a header information specifying the starting line number and the number of affected lines.
Using this information, we can get the range of the modification (for a given file pair: original and new).
To obtain a more accurate result, we subtracted the unmodified code lines from this range.
Although the diff files generated by GitHub contain additional information about which method is affected, it does not carry enough information because the difference can affect multiple source code elements (overlapping cases that are not handled by GitHub).
Thus, there is no further task but to examine the source code elements in every modified file and identify which ones of them are affected by the changes.
The method uses the source code element positions, i.e., source line mappings from the output of the OpenStaticAnalyzer tool.
We identified the source code elements by their fully qualified names that involve the name of the package, the class, the method, the type of the parameters, and the type of the return value.

Next, we take the commits that were selected by the ``Processing Raw Data'' step and mark the code sections affected by the bug in these commits.
We do this by accumulating the modifications on the issue level and collecting the fully qualified names of the elements.
Then, the algorithm marks the source code elements in the appropriate versions that will be entries in the dataset (touched in order to fix a bug).
If a source code element in a specific version is marked by multiple issues, then it contains multiple bugs in that version.
The dataset for files, classes, and methods are exported into three different files in a simple CSV format.
The first row of these files contains the header information, namely the commit id, the qualified name and the bug cardinality.
Further lines store the data of the source code elements according to the header.

\subsection{Combining CSV files}

Now, the CSV outputs of OpenStaticAnalyzer and the previously described CSV output can be merged.
In this phase, we attach the source code elements that are entries in the dataset to the calculated metrics.
The output of this step is also a CSV file for each type of source code element, containing the hash code of the version, unique identifiers of the source code elements, identifiers of metrics, rule violation groups, and bug cardinality (the number of bugs located in the source code elements).
One entry is equivalent to one source code element at a given time (the same source code element can occur more than once with a different commit id -- hash).

\subsection{Filtering}\label{sec:filtering}

This data set we compiled so far can contain various entries that complicate further investigations.
As the data set should be suitable for studying the connection between different metrics and bug occurrences, it should serve as a practical input for different machine learning algorithms.
It is possible, however, to have entries in the dataset that have the same metric values with different number of bugs assigned to them.
For example, let us consider a buggy method $f$ with metric values $M_{f_1}$.
After the bugfix, the metric values of $f$ is changed to $M_{f_2}$.
Similarly, let us consider another buggy method $g$ with metric values $M_{g_1}$ and $M_{g_2}$, respectively.
These two methods could contain two different bugs that are present in a system for distinct periods of time.
In this case, the dataset would contain 4 entries: $M_{f_1}$, $M_{f_2}$, $M_{g_1}$, $M_{g_2}$, where $M_{f_1}$ and $M_{g_1}$ are buggy and $M_{f_2}$ and $M_{g_2}$ are non-buggy entries.
If any of these metric values are equal (e.g. $M_{f_1}=M_{g_2}$ or $M_{g_1}=M_{g_2}$), then redundancy occurs that can influence the accuracy of machine learning for bug prediction (overfitting, contradicting records).

To solve this issue, we used different approaches to filter the raw dataset and eliminate the redundant entries.
We tried various methods to reduce the noise in the learning set, whose entries are classified into either buggy or not buggy.

\begin{itemize}
    \item \textbf{Removal}: keep the entries located in the class with the larger cardinality (e.g., for a 10:20 distribution, the result is 0:20)
    \item \textbf{Subtract}: reduce the number of entries in the class with the larger cardinality by removing as many entries as the cardinality of the smaller class (e.g., for a 10:20 distribution, the result is 0:10)
    \item \textbf{Single}: remove the entries of the class with the smaller cardinality and hold only one entry from the larger one (e.g., for a 10:20 distribution, the result is 0:1)
    \item \textbf{GCF}: divide the number of entries of both classes by their greatest common factor (or greatest common divisor) and retain only the resulting amounts of entries from the classes (e.g., for a 10:20 distribution, the result is 1:2)
\end{itemize}

Each selected approach can seriously modify the result set, thus we investigated all four options and additionally the basic case when no filtering was applied.
Tables~\ref{filteringresultsmethod},~\ref{filteringresultsclass}, and~\ref{filteringresultsfile} present average F-measure values calculated for all of the machine learning algorithms we used for all of the projects.
From these tables we can see that the Single and GCF methods performed quite similarly but were less effective than Subtract or Removal.

We employed a statistical significance test, namely the Friedman test~\cite{friedman1940comparison} with a threshold of $\alpha=0.05$ to assess the significance of the differences between the averages, as it was done similarly in previous bug prediction studies~\cite{herbold2018comparative, ghotra2015revisiting}.
Our data does not follow normal distribution, it consists of dependent samples and we have five paired groups, thus the Friedman test is the appropriate choice.
The null hypothesis is that the multiple paired samples have the same distribution.
The tests resulted in very low $p$ values ($p_{method}=5.32\text{e-}80$, $p_{class}=2.03\text{e-}77$, $p_{file}=1.83\text{e-}40$); therefore, we reject the null hypothesis which means the distributions are not equal.
Then, we applied the Nemenyi post-hoc test~\cite{nemenyi} ($\alpha=0.05$) that is usually used after a null hypothesis is rejected to gain more insight on the significance of the differences.
The critical value for 5 groups and 176 samples (11 machine learning algorithms $\times$ 16 databases) based on the studentized range table is $q_{crit}=3.9$.
Tables~\ref{filteringresultsmethod_sig},~\ref{filteringresultsclass_sig}, and~\ref{filteringresultsfile_sig} list the resulted $p$ values with the corresponding rank difference in parentheses.

Let us consider the method level F-measure values in Table~\ref{filteringresultsmethod} where Removal has the highest average F-measure (0.5773) and Subtract is a close second (0.5717).
In Table~\ref{filteringresultsmethod_sig}, the results of the significance tests for method level show that the $p$ value of the test between Subtract and No filter is below the threshold ($p=0.001<\alpha=0.05$); therefore, the difference is significant and with Subtract having a higher average F-measure (0.5717) than No filter (0.5317), we can state that it is significantly better.
We can conclude the same when comparing Subtract with Single ($p=0.001<\alpha=0.05$) or with GCF ($p=0.001<\alpha=0.05$).
The $p$ value between Subtract and Removal is $p=0.210>\alpha=0.05$ which is not significant.

Similar results can be concluded for class level and for file level as well.
We can state that the Removal and Subtract methods performed significantly better than the other methods in all three cases.
The difference between the Removal and Subtract methods is not significant.

We speculate that a disadvantage to Single is that it drops the multiplicity of the records (i.e., the weight information).
The problem with GCF, on the other hand, is that it will only perform filtering when the greatest common factor is not one, and that it does not eliminate the noise completely (i.e., it will keep at least one entry from both classes).
Removal removes the noise entirely, but it suffers from the fact that it ignores the minority.

The Subtract method, however, neutralizes the positive and negative entries with identical feature vectors.
This means that it removes the noise while also keeping the weight of the records, so this filtering method seems to be the best choice.
Presenting all the five different sets would be lengthy, thus we will only present the results achieved by the Subtract method.

\begin{table}[H]
\caption{Filtering results at method level}
\fontsize{10pt}{12pt}\selectfont
\def\arraystretch{1}
\centering
\setlength{\tabcolsep}{6pt}
\begin{tabular}{|l|c|c|c|}
\hline
\textbf{Method} & \textbf{Precision} & \textbf{Recall} & \textbf{F-Measure}  \\ \hline
No filter   & 0.5553 & 0.5501 & 0.5317  \\ 
Removal     & 0.6070 & 0.5963 & 0.5773 \\ 
Subtract    & 0.5974 & 0.5893 & 0.5717 \\ 
Single      & 0.5495 & 0.5448 & 0.5250 \\ 
GCF         & 0.5445 & 0.5408 & 0.5218 \\ \hline
\end{tabular}
\label{filteringresultsmethod}
\end{table}

\begin{table}[H]
\caption{Significance test results for method level filtering}
\fontsize{8pt}{10pt}\selectfont
\def\arraystretch{1}
\centering
\setlength{\tabcolsep}{6pt}
\begin{tabular}{|l|c|c|c|c|}
\hline
			& \multicolumn{1}{c|}{\textit{No filter}} & \multicolumn{1}{c|}{\textit{Removal}} & \multicolumn{1}{c|}{\textit{Subtract}} & \multicolumn{1}{c|}{\textit{Single}} \\
\hline
\textit{Removal} & \textbf{0.001} (16.8763) & \gcell & \gcell & \gcell \\
\hline
\textit{Subtract} & \textbf{0.001} (13.8729) & 0.210 (3.0034) & \gcell & \gcell  \\
\hline
\textit{Single} & 0.736 (1.6686) & \textbf{0.001} (18.5449) & \textbf{0.001} (15.5414) & \gcell  \\
\hline
\textit{GCF} & \textbf{0.020} (4.2906) & \textbf{0.001} (21.1669) & \textbf{0.001} (18.1635) & 0.343 (2.6220)  \\
\hline
\end{tabular}
\label{filteringresultsmethod_sig}
\end{table}

\begin{table}[H]
\caption{Filtering results at class level}
\fontsize{10pt}{12pt}\selectfont
\def\arraystretch{1}
\centering
\setlength{\tabcolsep}{6pt}
\begin{tabular}{|l|c|c|c|}
\hline
\textbf{Method} & \textbf{Precision} & \textbf{Recall} & \textbf{F-Measure}  \\ \hline
No filter   & 0.5265 & 0.5235 & 0.5128 \\ 
Removal     & 0.5567 & 0.5528 & 0.5419 \\ 
Subtract    & 0.5541 & 0.5499 & 0.5393 \\ 
Single      & 0.5236 & 0.5206 & 0.5090 \\ 
GCF         & 0.5221 & 0.5201 & 0.5077 \\ \hline
\end{tabular}
\label{filteringresultsclass}
\end{table}

\begin{table}[H]
\caption{Significance test results for class level filtering}
\fontsize{8pt}{10pt}\selectfont
\def\arraystretch{1}
\centering
\setlength{\tabcolsep}{6pt}
    \begin{tabular}{|l|c|c|c|c|}
    \hline
          & \multicolumn{1}{c|}{\textit{No filter}} & \multicolumn{1}{c|}{\textit{Removal}} & \multicolumn{1}{c|}{\textit{Subtract}} & \multicolumn{1}{c|}{\textit{Single}}  \\
    \hline
    \textit{Removal} & \textbf{0.0010} (15.9228) & \gcell & \gcell & \gcell \\
    \hline
    \textit{Subtract} & \textbf{0.0010} (15.3984) & 0.9000 (0.5244) & \gcell & \gcell \\
    \hline
    \textit{Single} & 0.9000 (0.7628) & \textbf{0.0010} (16.6856) & \textbf{0.0010} (16.1612) & \gcell \\
    \hline
    \textit{GCF} & \textbf{0.0496} (3.8615) & \textbf{0.0010} (19.7844) & \textbf{0.0010} (19.2599) & 0.1828 (3.0988) \\
    \hline
    \end{tabular}
\label{filteringresultsclass_sig}
\end{table}

\begin{table}[H]
\caption{Filtering results at file level}
\fontsize{10pt}{12pt}\selectfont
\def\arraystretch{1}
\centering
\setlength{\tabcolsep}{6pt}
\begin{tabular}{|l|c|c|c|}
\hline
\textbf{Method} & \textbf{Precision} & \textbf{Recall} & \textbf{F-Measure}  \\ \hline
No filter   & 0.5160 & 0.5117 & 0.4883  \\ 
Removal     & 0.5451 & 0.5414 & 0.5194 \\ 
Subtract    & 0.5407 & 0.5371 & 0.5147 \\ 
Single      & 0.5187 & 0.5148 & 0.4910 \\ 
GCF         & 0.5172 & 0.5129 & 0.4889 \\ \hline
\end{tabular}
\label{filteringresultsfile}
\end{table}

\begin{table}[H]
\caption{Significance test results for file level filtering}
\fontsize{8pt}{10pt}\selectfont
\def\arraystretch{1}
\centering
\setlength{\tabcolsep}{6pt}
    \begin{tabular}{|l|c|c|c|c|}
    \hline
          & \multicolumn{1}{c|}{\textit{No filter}} & \multicolumn{1}{c|}{\textit{Removal}} & \multicolumn{1}{c|}{\textit{Subtract}} & \multicolumn{1}{c|}{\textit{Single}} \\
    \hline
    \textit{Removal} & \textbf{0.0010} (14.2066) & \gcell & \gcell & \gcell \\
    \hline
    \textit{Subtract} & \textbf{0.0010} (13.2055) & 0.9000 (1.0011) & \gcell & \gcell \\
    \hline
    \textit{Single} & 0.0682 (3.6947) & \textbf{0.0010} (10.5119) & \textbf{0.0010} (9.5107) & \gcell \\
    \hline
    \textit{GCF} & 0.9000 (0.3576) & \textbf{0.0010} (13.8490) & \textbf{0.0010} (12.8479) & 0.1262 (3.3371) \\
    \hline
    \end{tabular}
\label{filteringresultsfile_sig}
\end{table}

\subsection{Classification and Resampling}\label{sec:resampling}

As trying to predict the exact number of bugs in a given source code element would be much more difficult -- and would presumably require much larger datasets -- we chose to restrict our study to predicting a boolean ``flag'' for the presence of any bugs.
Thus, we applied only classification algorithms, and to do so, classes need to be formed from the bug numbers.
For the binary classification, we selected instances with zero bugs into one class (non-buggy), and the remaining ones~--~with one or more bugs~--~into the second class (buggy).

Another problem we faced is that imbalanced learning sets could be formed from the dataset, where the positive or negative entries are in a majority which could also be misleading for model training.
For example, the ratio of buggy and non-buggy source code elements or files can be totally different.

To handle this issue at the machine learning level, we used random under sampling~\cite{he2009learning,wang2013using} to obtain an equivalent number of elements in the two categories.
For instance, if we have a final set as a corpus at method level that contains 10 buggy methods and 50 non-buggy methods, we use random under sampling for the non-buggy set to decrease the number of samples and balance the ratio to 10-10.
The training process -- using this random under sampling -- is repeated multiple times and finally, an average is calculated.
Without random under sampling, the machine learning algorithms achieved very high precision, recall, and F-measure values (e.g., by classifying all elements as non-buggy) because a significantly large difference was usually present in the number of entries for the two classes (non-buggy vs buggy).
For this reason, we present only the results achieved by using random under sampling.

\subsection{The BugHunter Dataset}\label{sec:dataset}

As a result and a main contribution, we constructed a novel kind of bug dataset that contains before/after fix states of source code elements at file, class, and method levels.
We produced a dataset for every project (see Table~\ref{tab:projects}) and also a combined one with all the projects included.
In Table~\ref{tab:metadata}, we collected the general metadata about the dataset which is scattered throughout in the paper.
The \emph{Ratio of the faulty entries} varies between the different granularity levels due to the nature of the bug-fixing changes.
For example, on method level it is more common to split a method into multiple parts or to introduce new methods to the source code during a fix, which results in more non-buggy entries than buggy, hence the low ratio.
On class level, however, the ratio is closer to 1.0 because it is less likely to create a new class in order to fix a bug.
Furthermore, on class and file levels, a new factor also contributes to the ratio.
Because classes and files represent a larger part of the source code than methods, it happens that in a fixed state of a bug the containing class or file contains other bugs as well.
In these cases, the entries related to the fixed state are marked as faulty in the dataset, which results in higher ratio values, above 1.0 in the case of files.

The resulting BugHunter Dataset 1.0 is available as an online appendix at:

\noindent
\url{http://www.inf.u-szeged.hu/~ferenc/papers/BugHunterDataSet/}
or

\noindent
\url{http://dx.doi.org/10.17632/8tx7kjbkg4.2}

\medskip
\noindent
The \texttt{BugHunterDataset-1.0.zip} file contains the dataset in CSV format as described above.
The directory named \texttt{full} contains the unfiltered database.
The remaining four directories, namely \texttt{gcf}, \texttt{remove}, \texttt{single}, and \texttt{subtract} contain the results of the different filtering methods.
Each of these directories contain 15 subdirectories -- one for each subject system -- and an additional directory named \texttt{all} which contains the summarized dataset.
Three CSV files are placed in these directories for file, class, and method levels, respectively.
There is also a fourth CSV file in each directory, called \texttt{method-p.csv} which contains the method level dataset extended with an additional column, the name of the parent class (see Section~\ref{sec:rq2}).
Additionally, the \texttt{appendix.zip} file contains the analysis results presented in Section~\ref{evaluation} in spreadsheet files.

\begin{table}[H]
\caption{BugHunter Dataset metadata}
\fontsize{8pt}{10pt}\selectfont
\def\arraystretch{1}
\centering
\setlength{\tabcolsep}{6pt}
    \begin{tabular}{|l|C{4,5cm}|}
    \hline
    \textbf{Type} & bug prediction dataset\\
    \hline
    \textbf{Granularity} & file, class, method \\
    \hline
    \textbf{Number of projects} & 15 \\
    \hline
    \textbf{Number of metrics} & static source code metrics (66), \newline code duplication metrics (8), \newline code smell metrics (35)  \\
    \hline
    \textbf{Number of entries} &  file: 70,088 \newline class: 84,562 \newline method: 159,078 \\
    \hline
    \textbf{Ratio of the faulty entries} & file: 1.03 (35,507/34,581) \newline class: 0.95 (41,098/43,475) \newline method: 0.59 (58,810/100,268) \\
    \hline
    \end{tabular}
\label{tab:metadata}
\end{table}

\subsection{Validation} \label{validation}

When constructing a dataset in an automatic way, one always has to validate the constructed set.
As seen previously, this kind of generated data should always be handled with mistrust~\cite{githubpromises}.
We chose the mid-sized project JUnit for such manual validation, which contains 90 bug reports (6 open and 84 closed) and a total of 72 referencing commits to the bugs, thus this project seems to be suitable for manual validation investing a reasonable amount of effort.
We validated the 84 closed bug reports manually to verify whether the bugs are valid, and whether the matching algorithm works well.
Table~\ref{validationresults} summarizes our findings.

\begin{table}[htb!]
\centering
\begin{tabular}{|c|c|c|c|c|}
\hline
Closed bugs & Bugs in dataset & Commits & Java code & Commit mismatch \\ \hline
84          & 37               & 72      & 61        & 5               \\ \hline
\end{tabular}
\caption{Validation results}
\label{validationresults}
\end{table}

From the total number of closed bugs, only 37 are present in the dataset because many fixing commits are not related to the source code (e.g., documentation) or Java code (e.g., bug in build XML).
This is shown in the Java code column of the table that summarizes the number of commits that contain Java language code (61).
These commits that include code are referencing 37 bugs, thus at least one bug exists that is referenced from multiple commits according to the pigeonhole principle.

We found 5 ``commit mismatch'' cases in total, where only comments were modified in the source code -- this means 7 entries in the dataset.
The dataset created for JUnit has 734 entries in total (92 files, 216 classes and 426 methods), thus a very small (0.95\%) number of entries was incorrectly included.
Out of the 734 entries, 286 are not related to test code (43 files, 77 classes, and 166 methods).
Based on this, we can presume that our validated dataset can be an appropriate corpus for further investigations and that our bug extraction mechanism is working quite reliably.

\subsection{Relationship of files and classes}

\begin{figure}[htb!]
\centering
\includegraphics[width=\textwidth]{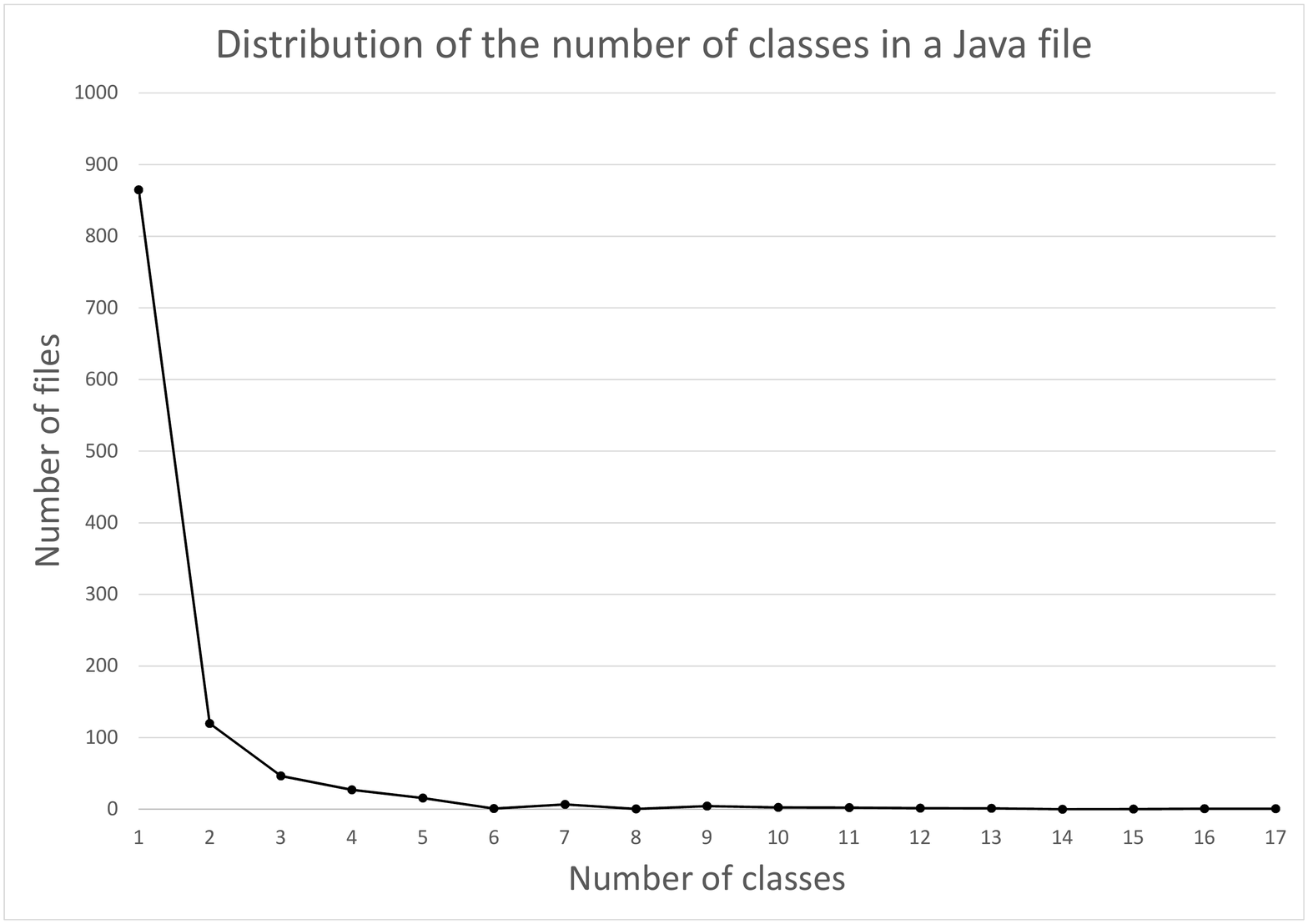}
\caption{Distribution of the number of classes in a Java file} 
\label{fig:distofclassesinfiles}
\end{figure}

In case of Java, there is usually one class per \textit{.java} file.
We examined how true this is for our subject projects.
We randomly chose 100 commits that we analyzed from each project and we counted the number of classes in each file, not including test files.
After we calculated the frequency of these values for each commit, we calculated the average frequency (see Figure~\ref{fig:distofclassesinfiles}).
The diagram shows that most of the files contain only a single class (865), but there is a significant number of files with more than one class (120 files with 2 classes, 46 files with 3 classes, etc.).

Although we have a larger set of metrics on class level than on file level, we cannot associate a file to a single class due to the one-to-many relationship between them.

\subsection{Computational cost of extending the dataset}

Extending the dataset comes with computational cost that depends on multiple factors.
Adding a new project to the dataset requires the project to have bug reports with bug-fixing commits.
Finding such projects is time-consuming, because it mostly requires manual work to select good candidates (see Section~\ref{sec:chosenprojects}).

The most critical step is to collect appropriate bugs.
For this initial dataset, we collected projects from GitHub, since its API makes it easy to gather the required information about bugs automatically.
The actual run-time of this step depends on the size of the project, e.g. number of commits and number of bug reports, but for the selected 15 projects, it took just a few hours to save the required data.
Selecting bugs manually would take considerably more time.
GitHub has a limit on the number of API requests per hour which increased the total run-time.
It is possible to collect data from other sources, although it may require a different amount of work.

The most time and resource consuming task is the source code analysis.
We used the OpenStaticAnalyzer tool which performs deep static analysis; therefore, it requires more resources than a simple parser tool.
During this step, we extract the static source code metrics and the source code positions of the classes and methods, as described in Sections~\ref{sec:processingrawdata} and \ref{sec:sourcecodeanalysis}, respectively.
The computational cost of this step highly depends on the number of bug-fixing commits, the size of the source code, and the analyzer tool.
It took days to analyze each of the nearly 10,000 bug-fixing commits.
There are other tools that could be used to extract the source code positions and other tools to compute metrics with potentially less run-time, but the tool we used produces a wide range of metrics and rule violations accurately in a well processable format.

The next step, determining the buggy source code elements, is a simple algorithm that does not require much resources.
The run-time here mostly depends on the number of bug-fixing commits.
It took only a few hours for the 15 projects.

For example, to process a smaller project such as jUnit, it took around 2 hours of machine time: 10 minutes to download the data from GitHub, 110 minutes to analyze 107 versions of the project (on average 1 minute per version) and around 2 minutes to produce the bug dataset entries.
Regarding a larger project, Elasticsearch, it took around 6 hours to download the data from GitHub, around 1,600 hours to analyze 4,881 versions of the project (on average 20 minutes per version) and it took around 90 minutes to produce the bug dataset.

At this point, the data is ready to be added to the dataset.
The last step is to match the format of the dataset (see Section~\ref{sec:dataset}).
Since the dataset consist of CSV files, it is very easy to extend it with new projects or with additional bugs for the projects that are already present.

\section{Evaluation} \label{evaluation}

To evaluate the usefulness of our new bug dataset, we created bug prediction models by using machine learning algorithms.
During the training, we used 10-fold cross validation to measure the accuracy of the models.
To compare the models, we used precision, recall, and F-measure metrics that are defined as follows:

\bigskip
\noindent
\begin{center}
$precision = \dfrac{TP}{TP + FP}$ \hspace{50pt}
$recall = \dfrac{TP}{TP + FN} $ \\ 	
\end{center}

\begin{center}
		$F-measure = 2\cdot \dfrac{ precision \cdot recall}{precision + recall}$, \\
\end{center}

\noindent where \textit{TP} (True Positive) is the number of methods/classes/files that were predicted as faulty and observed as faulty, \textit{FP} (False Positive) is the number of methods/classes/files that were predicted as faulty but observed as not faulty, \textit{FN} (False Negative) is the number of methods/classes/files that were predicted as non-faulty but observed as faulty.
We used the following algorithms from the Weka library to find out how they perform on our dataset:

\begin{itemize}
	\setlength\itemsep{-0.3em}
		\item NaiveBayes
		\item NaiveBayesMultinomial
		\item Logistic
		\item SGD
		\item SimpleLogistic
		\item VotedPerceptron
		\item DecisionTable
		\item OneR
		\item J48 (C4.5)
		\item RandomForest
		\item RandomTree
\end{itemize}

We have 3 source code levels (method, class, file), 15 chosen projects (plus the summarized dataset), and 11 machine learning algorithms, implying that full tables with the obtained results would be too large to present in the paper, thus we introduce only the best algorithms here for the overall dataset.
Please note that the online appendix (see Section~\ref{sec:dataset} for the Web link) contains all the analysis results in spreadsheet files.

\subsection{First Research Question}\label{sec:rq1}

The first research question we will answer is the following:

\bigskip
\fbox{
  \parbox{\textwidth - 45pt}{
    \textbf{RQ1:} \textit{Is the constructed dataset usable for bug prediction purposes?}
  }
}
\bigskip

To answer RQ1, we present the best results obtained by different machine learning algorithms at method, class, and file level.
Similar to Section~\ref{sec:filtering}, we used the Friedman test and the Nemenyi post-hoc test to check whether the distributions of the samples are equal or not.
We observed the same, very low $p$ values ($p_{method}=1.21\text{e-}14$, $p_{class}=1.54\text{e-}07$, $p_{projected}=3.77\text{e-}14$, $p_{file}=1.06\text{e-}05$), thus the distributions are not equal.
With $\alpha=0.05$, the critical value in this case for 11 groups (machine learning algorithms) and 16 samples (databases) based on the studentized range table is $q_{crit}=5.256$.
Due to the size of the tables, we do not include them here, instead, the complete tables are available in the Appendix (see Tables~\ref{tab:sign_res_method_ml},~\ref{tab:sign_res_class_ml},~\ref{tab:sign_res_projected_ml}, and~\ref{tab:sign_res_file_ml}).

\subsubsection{Method level}

We trained models to use method level metrics to predict future failures at method level.
The results are shown in Table~\ref{methodtop5} containing the best five algorithms selected by F-measure values.

The fifth best algorithm with 0.5983 F-measure value is DecisionTable~\cite{Kohavi1995}.
The SimpleLogistic algorithm resulted a slightly higher F-measure (0.6031).
SimpleLogistic algorithm builds linear logistic regression models that uses automatic attribute selection~\cite{Landwehr2005}.
The first three algorithms are all from the tree family.
J48~\cite{Quinlan1993} that uses pruned or unpruned C4.5 decision tree to build a model is the second best algorithm (0.6119).
The third and the first algorithms also use trees to produce prediction models.
RandomForest (0.6319) builds a forest from RandomTrees to get a slightly better result than RandomTree (0.6110).
The results of the statistical tests in Table~\ref{tab:sign_res_method_ml} show that the differences between the top five algorithms are not statistically significant ($p>\alpha=0.05$) but the difference between the worst (NaiveBayes, NaiveBayesMultinomial and VotedPerceptron) and the best performing algorithms are significant.

\begin{table}[H]
\caption{TOP 5 machine learning algorithms for method level based on F-measure}
\fontsize{10pt}{12pt}\selectfont
\def\arraystretch{1}
\centering
\setlength{\tabcolsep}{6pt}
\begin{tabular}{|l|c|c|c|}
\hline
\textbf{Algorithm}         & \textbf{Precision} & \textbf{Recall} & \textbf{F-Measure}  \\ \hline
trees.RandomForest         & 0.6335 &	0.6324 & 0.6319 \\
trees.J48                  & 0.6147 &	0.6134 & 0.6119 \\
trees.RandomTree           & 0.6115 &	0.6113 & 0.6110 \\
functions.SimpleLogistic   & 0.6062 &	0.6043 & 0.6031 \\
rules.DecisionTable        & 0.6138 &	0.6073 & 0.5983 \\ \hline
\end{tabular}
\label{methodtop5}
\end{table}

At method level, trees were performing the best and can result up to 0.6319 when considering F-measure values.
We also investigated the results by projects and found that specific projects performed worse than others.
Android Universal Image Loader and Ceylon were the worst, both when considering precision, recall, or F-measure.
Achieved F-measure values depend highly upon the project itself.
A possible factor that plays role in this is the size of the built corpus.
These projects have a smaller training corpus and more inconsistency in the feature vectors, consequently it is harder to build a well-performing prediction model.
This phenomenon does not appear only at method level but at class and file level too, since at these levels even less entries are created in the dataset.
The best F-measure values (over 0.75 in one case) achieved on different projects are demonstrated in Table~\ref{tab:bestbyproject_method}.

\begin{table}[H]
\caption{The best F-measure values by projects at method level}
\fontsize{10pt}{12pt}\selectfont
\def\arraystretch{1}
\centering
\setlength{\tabcolsep}{6pt}
\begin{tabular}{|l|c|c|}
\hline
\textbf{Project}                   & \textbf{F-measure} & \textbf{Algorithm}        \\ \hline
antlr4                             & 0.7573             & trees.RandomForest        \\
BroadleafCommerce                  & 0.7366             & trees.RandomForest        \\
hazelcast                          & 0.7170             & trees.RandomForest        \\
mct                                & 0.6876             & trees.RandomForest        \\
oryx                               & 0.6678             & trees.RandomForest        \\
junit                              & 0.6638             & rules.DecisionTable       \\
all                                & 0.6622             & trees.RandomForest        \\
netty                              & 0.6412             & trees.RandomForest        \\
elasticsearch                      & 0.6411             & trees.RandomForest        \\
orientdb                           & 0.6236             & trees.RandomForest        \\
titan                              & 0.6216             & functions.SGD             \\
neo4j                              & 0.6086             & functions.Logistic        \\
mcMMO                              & 0.5815             & trees.RandomForest        \\
MapDB                              & 0.5610             & functions.SimpleLogistic  \\
Android-Universal-Image-Loader     & 0.5569             & functions.SGD             \\
ceylon-ide-eclipse                 & 0.5395             & trees.RandomTree          \\ \hline
\end{tabular}
\label{tab:bestbyproject_method}
\end{table}

\subsubsection{Class level}

When considering class level, we have quite a different set of algorithms in the top five than in the case of methods.
Furthermore, the precision, recall, and F-measure values differ significantly from those we obtained at method level.
We suspect that the main reason behind this is the different set of metrics used to build models, thus to predict the possibility of occurring bugs in a class.
At class level, simple logistic, decision table, and SGD were the best.
Function and rule based groups of machine learning algorithms can be emphasized as the best when considering class level.
The best machine learning algorithms at class level are shown in Table~\ref{fmeasureclasstop5} with F-measure values around 0.56.
In Table~\ref{tab:sign_res_class_ml}, the results of the significance tests show that the best algorithm, SimpleLogistic with 0.5685 F-measure, achieved significantly better results than the worst two algorithms that are not in the top five (NaiveBayes $p=0.019$ and VotedPerceptron $p=0.001$).
Between the top five algorithms, the differences are not significant ($p>\alpha=0.05$).

The low F-measures values suggest that one cannot build efficient prediction models at class level.

\begin{table}[H]
\caption{TOP 5 machine learning algorithms for class level based on F-measure}
\fontsize{10pt}{12pt}\selectfont
\def\arraystretch{1}
\centering
\setlength{\tabcolsep}{6pt}
\begin{tabular}{|l|c|c|c|}
\hline
\textbf{Algorithm} & \textbf{Precision} & \textbf{Recall} & \textbf{F-Measure}  \\ \hline
functions.SimpleLogistic & 0.5760 & 0.5763 &  0.5685 \\
rules.DecisionTable      & 0.5703 & 0.5705 &  0.5637 \\
functions.SGD            & 0.5718 & 0.5676 &  0.5626 \\
functions.Logistic       & 0.5561 & 0.5552 &  0.5537 \\
trees.J48                & 0.5531 & 0.5530 &  0.5520 \\ \hline
\end{tabular}
\label{fmeasureclasstop5}
\end{table}

\begin{table}[H]
\caption{The best F-measure values by projects at class level}
\fontsize{10pt}{12pt}\selectfont
\def\arraystretch{1}
\centering
\setlength{\tabcolsep}{6pt}
\begin{tabular}{|l|c|c|}
\hline
\textbf{Project}                   & \textbf{F-measure} & \textbf{Algorithm}        \\ \hline
BroadleafCommerce                  & 0.7400             & trees.RandomForest        \\
oryx                               & 0.7095             & functions.SGD             \\
junit                              & 0.6639             & rules.DecisionTable       \\
hazelcast                          & 0.6175             & trees.RandomTree          \\
MapDB                              & 0.6138             & functions.SimpleLogistic  \\
orientdb                           & 0.6132             & functions.SimpleLogistic  \\
mct                                & 0.5825             & functions.SGD             \\
elasticsearch                      & 0.5817             & rules.DecisionTable       \\
all                                & 0.5803             & rules.DecisionTable       \\
ceylon-ide-eclipse                 & 0.5789             & rules.DecisionTable       \\
antlr4                             & 0.5685             & rules.OneR                \\
mcMMO                              & 0.5670             & functions.SGD             \\
titan                              & 0.5614             & functions.SimpleLogistic  \\
netty                              & 0.5537             & functions.Logistic        \\
neo4j                              & 0.5413             & functions.SimpleLogistic  \\
Android-Universal-Image-Loader     & 0.4713             & bayes.NaiveBayes          \\ \hline
\end{tabular}
\label{tab:bestbyproject_class}
\end{table}

However, we present the F-measure values of individual projects in Table~\ref{tab:bestbyproject_class}.
Considering these F-measure values, we can see the same phenomenon as in the case of methods.
McMMO, Android Universal Image Loader, and Neo4J are in the worst 5, which supports the previous experience according to which different projects provide different amount of ``munition'' for predicting faults.
The best case, however, provides an F-measure of 0.74.

\subsubsection{File level}

In Java context, a public class is almost equivalent to a file with a \textit{'.java'} extension.
However, despite the fact that we compute a different set of metrics for class and file level, the results are quite similar.
Since we operate on a different set of metrics at class and file level, this explains that different machine learning algorithms performed the best.
The best algorithms for this level use tree-based approaches to predict bugs as it is shown in Table~\ref{fmeasurefiletop5}.
Similar to class level, the differences between the top five algorithms are not considered significant ($p>\alpha=0.05$), as can be seen in Table~\ref{tab:sign_res_file_ml}.
The best performing algorithm, achieving an F-measure value of 0.5476, is RandomTree.
The top five algorithms achieved significantly better results compared to the worst algorithm (VotedPerceptron).
Between the top five algorithms, the differences are not considered significant ($p>\alpha=0.05$).

\begin{table}[H]
\caption{TOP 5 machine learning algorithms for file level based on F-measure}
\fontsize{10pt}{12pt}\selectfont
\def\arraystretch{1}
\centering
\setlength{\tabcolsep}{6pt}
\begin{tabular}{|l|c|c|c|}
\hline
\textbf{Algorithm}  & \textbf{Precision} & \textbf{Recall} & \textbf{F-Measure}  \\ \hline
trees.RandomTree          & 0.5484 & 0.5484 &	0.5476 \\
trees.RandomForest        & 0.5458 & 0.5461 &	0.5455 \\
functions.Logistic        & 0.5528 & 0.5474 &	0.5367 \\
rules.OneR                & 0.5358 & 0.5359 &	0.5348 \\
functions.SimpleLogistic  & 0.5491 & 0.5474 &	0.5321 \\ \hline
\end{tabular}
\label{fmeasurefiletop5}
\end{table}

We also present the F-measure values obtained on projects in Table~\ref{tab:bestbyproject_file}.
The takeaway remains the same, Android Universal Image Loader and Neo4J are located in the worst five projects again.
On the other hand, Broadleaf Commerce, Oryx, and Hazelcast seem to be appropriate to use in model building.
The best F-measure value is over 0.77.

\begin{table}[H]
\caption{The best F-measure values by projects at file level}
\fontsize{10pt}{12pt}\selectfont
\def\arraystretch{1}
\centering
\setlength{\tabcolsep}{6pt}
\resizebox{\textwidth}{!}{
\begin{tabular}{|l|c|c|}
\hline
\textbf{Project}                   & \textbf{F-measure} & \textbf{Algorithm}          \\ \hline
BroadleafCommerce                  & 0.7741             & trees.RandomForest          \\
oryx                               & 0.6458             & bayes.NaiveBayesMultinomial \\
hazelcast                          & 0.6417             & trees.RandomTree            \\
all                                & 0.6234             & trees.RandomTree            \\
orientdb                           & 0.6200             & rules.DecisionTable         \\
elasticsearch                      & 0.6073             & trees.RandomTree            \\
ceylon-ide-eclipse                 & 0.5857             & trees.J48                   \\
titan                              & 0.5793             & functions.SimpleLogistic    \\
mcMMO                              & 0.5702             & trees.RandomForest          \\
MapDB                              & 0.5525             & rules.OneR                  \\
junit                              & 0.5484             & rules.OneR                  \\
netty                              & 0.5344             & trees.RandomTree            \\
antlr4                             & 0.5212             & trees.RandomForest          \\
neo4j                              & 0.5138             & rules.DecisionTable         \\
Android-Universal-Image-Loader     & 0.4781             & rules.DecisionTable         \\
mct                                & 0.4576             & functions.Logistic          \\ \hline
\end{tabular}
}
\label{tab:bestbyproject_file}
\end{table}

\vspace{6pt}
\setlength\parindent{0pt}
\resizebox{\textwidth}{!}{
\fbox{
  \parbox{\textwidth - 15pt}{
		\textbf{Answering RQ1:} \textit{
Considering the results we obtained, we can state that creating bug prediction models at method level is more successful than at file and class levels if we consider the full dataset.
We also showed the diversion in F-measure values by projects, which strengthens our assumption that not all projects are capable of providing an appropriate training set.
We can obtain F-measure values on separate projects up to 0.7573, 0.7400, and 0.7741 at method, class and file level, respectively, which is promising. 
In our next research question, we accomplish an experiment and its results are even better.
However, even without knowing that there is a better solution, we can answer this research question in a positive manner and say that the constructed dataset is usable for bug prediction.
    }
  }
}
}
\vspace{6pt}
\setlength\parindent{15pt}

\subsection{Second Research Question}\label{sec:rq2}

The dataset contains the bug information on both method and class levels, and we also know the containing relationships between classes and methods.
However, since classes have a different source code metrics set than methods, a question arose: can we (and more importantly, should we) use method level metrics to predict faulty classes?
The second research question we will answer is the following:

\vspace{6pt}
\setlength\parindent{0pt}
\fbox{
  \parbox{\textwidth - 15pt}{
		\textbf{RQ2:} \textit{Are the method level metrics projected to class level better predictors than the class level metrics themselves?}
  }
}
\vspace{6pt}
\setlength\parindent{15pt}

We carried out an experiment where we projected the results of the method-level learning to the class level.
During the cross-validation of the method level learning, we used the containing classes of the methods to calculate the confusion matrix from the number of classes classified as buggy and non-buggy.
Classes containing at least one buggy method were considered as buggy.

We compared this result with the result of the class-level prediction.
The results in Table~\ref{aggregatedresults} show that the projection method performs much better than the prediction with class level metrics.

We applied the Wilcoxon-signed-rank test~\cite{wilcoxon1945individual} (a non-parametric paired test for dependent samples), with a threshold of $Z_{crit}=1.96$ (for a two-tailed test with $\alpha=0.05$) to check whether the difference is significant.
We also calculated the effect size of these tests with the Pearson correlation coefficient (Pearson's r) from the formula $r=\frac{Z}{\sqrt{N}}$, where $N$ is the total number of samples and $Z$ is the z-score of the test~\cite{rosenthal1994parametric}.
According to Cohen~\cite{cohen1992power}, the effect size is considered small if $r\approx0.1$, medium if $r\approx0.3$ or large if $r\approx0.5$.

After the test, we can confirm that the difference between the projection method and the prediction with class level metrics is significant ($Z=10.9>Z_{crit}=1.96$) and the effect size is considered large ($r=0.58$).

We suspect that this is due to the generality of class-level metrics, which are therefore not powerful enough to effectively distinguish source code bugs.
Although the bug information for methods does not include all bugs that affect the containing class (e.g. change of fields, interfaces or superclasses), method level metrics are more useful for bug prediction.

The results of the significance tests between the different machine learning algorithms is displayed in Table~\ref{tab:sign_res_projected_ml}.
The best performing algorithm is RandomForest with 0.7405 F-measure and it is significantly better than the worst three algorithms that are not displayed in Table~\ref{aggregatedresults} (NaiveBayes $p=0.001$; NaiveBayesMultinomial $p=0.001$; VotedPerceptron $p=0.003$).
The difference between the top five algorithms is not considered significant ($p>\alpha=0.05$).

\begin{table}[H]
\caption{The results of projected learning}
\fontsize{8pt}{10pt}\selectfont
\def\arraystretch{1}
\centering
\setlength{\tabcolsep}{3pt}
\begin{tabular}{|c|c|c|c|c|c|c|}
\hline
\multirow{2}{*}{\textbf{Algorithm}} & \multicolumn{2}{|c|}{\textbf{Precision}} & \multicolumn{2}{|c|}{\textbf{Recall}} & \multicolumn{2}{|c|}{\textbf{F-Measure}}  \\ \cline{2-7}
 & Projected & Class & Projected & Class & Projected & Class  \\ \hline
trees.RandomForest   & 0.7471 & 0.5336 & 0.7370 & 0.5336 & 0.7405 & 0.5334 \\ \hline
trees.RandomTree     & 0.7421 & 0.5381 & 0.7273 & 0.5380 & 0.7330 & 0.5376 \\ \hline
functions.SGD        & 0.7441 & 0.5718 & 0.7288 & 0.5676 & 0.7322 & 0.5626 \\ \hline
rules.DecisionTable  & 0.7425 & 0.5703 & 0.7404 & 0.5705 & 0.7309 & 0.5637 \\ \hline
trees.J48            & 0.7390 & 0.5531 & 0.7250 & 0.5530 & 0.7290 & 0.5520 \\ \hline
\end{tabular}
\label{aggregatedresults}
\end{table}

When using the projection approach to predict bugs in classes, the F-measure values reach 0.74.
As an extension of the answer to RQ1, we can provide the above described mechanism to locate class level bugs with a higher accuracy in a software system.

\vspace{6pt}
\setlength\parindent{0pt}
\fbox{
  \parbox{\textwidth - 15pt}{
		\textbf{Answering RQ2:} \textit{
		Using method level metrics for class level bug prediction performed the best in our study.
		This fact also contributes to the answer given for RQ1.
		Furthermore, method level metrics are better predictors when projected to class level than class level metrics by themselves.
		}
  }
}
\vspace{6pt}
\setlength\parindent{15pt}

\subsection{Third Research Question}\label{sec:rq3}

The third research question we will answer is the following:

\vspace{6pt}
\setlength\parindent{0pt}
\fbox{
  \parbox{\textwidth - 15pt}{
	\textbf{RQ3:} Is the \emph{BugHunter Dataset} more powerful and expressive than the \emph{GitHub Bug Dataset}, which is constructed with the traditional approach?
  }
}
\vspace{6pt}
\setlength\parindent{15pt}

Comparing the expressive power of different datasets is a harsh task since the various datasets were created with different purposes, they often have only few independent variables in common.
The projects included in these datasets are different as well.
Therefore, we provide an objective comparison between our previously published traditional bug dataset, the \emph{GitHub Bug Dataset}~\cite{toth2016public} and the \emph{BugHunter Dataset} in the following.
These two datasets include exactly the same 15 projects and the set of independent variables are common and also calculated in the same way with the same tool.
We used the same machine learning algorithms to build prediction models.
This way, it is quite straightforward to compare the expressiveness and compactness of these datasets.

\begin{table}[H]
\caption{Comparison of the size of the datasets}
\fontsize{8pt}{10pt}\selectfont
\def\arraystretch{1}
\centering
\setlength{\tabcolsep}{3pt}
\begin{tabular}{|c|r|r|r|r|r|r|r|r|r|}
\hline
\multirow{2}{*}{\textbf{Project}} & \multicolumn{3}{c|}{\textbf{Method}} & \multicolumn{3}{c|}{\textbf{Class}} & \multicolumn{3}{c|}{\textbf{File}}  \\ \cline{2-10}
                     & Trad       & BH     & Rate   & Trad        & BH     & Rate  & Trad        & BH     & Rate  \\ \hline
Android-U.-I.-L.     & 432        & 325    & 1.33   & 73          & 156    & 0.47  & 63          & 145    & 0.43  \\ 
antlr4               & 3,640      & 840    & 4.33   & 479         & 314    & 1.53  & 411         & 347    & 1.18  \\ 
BroadleafCommerce    & 14,651     & 4,709  & 3.11   & 1,593       & 2,957  & 0.54  & 1,719       & 2,969  & 0.58  \\ 
ceylon-ide-eclipse   & 8,787      & 2,087  & 4.21   & 1,611       & 1,275  & 1.26  & 700         & 946    & 0.74  \\ 
elasticsearch        & 34,324     & 35,862 & 0.96   & 5,908       & 24,994 & 0.24  & 3,035       & 17,724 & 0.17  \\ 
hazelcast            & 21,642     & 32,973 & 0.66   & 3,412       & 19,845 & 0.17  & 2,228       & 14,913 & 0.15  \\ 
junit                & 2,441      & 462    & 5.28   & 731         & 316    & 2.31  & 309         & 177    & 1.75  \\ 
MapDB                & 2,913      & 1,456  & 2.00   & 331         & 899    & 0.37  & 138         & 482    & 0.29  \\ 
mcMMO                & 2,531      & 1,184  & 2.14   & 301         & 732    & 0.41  & 267         & 678    & 0.39  \\ 
mct                  & 9,836      & 105    & 93.68  & 1,887       & 66     & 28.59 & 413         & 52     & 7.94  \\ 
neo4j                & 30,256     & 7,030  & 4.30   & 5,899       & 3,701  & 1.59  & 3,278       & 2,934  & 1.12  \\ 
netty                & 8,312      & 11,171 & 0.74   & 1,143       & 5,677  & 0.20  & 914         & 4,023  & 0.23  \\ 
orientdb             & 17,013     & 9,445  & 1.80   & 1,847       & 4,134  & 0.45  & 1,503       & 3,564  & 0.42  \\ 
oryx                 & 2,506      & 810    & 3.09   & 533         & 598    & 0.89  & 281         & 536    & 0.52  \\ 
titan                & 8,424      & 785    & 10.73  & 1,468       & 428    & 3.43  & 976         & 378    & 2.58  \\ \hline \hline
\textbf{Total}       & 167,708    & 109,244& \textbf{1.54} & 27,216 & 66,092 & \textbf{0.41} & 16,235 & 49,868 & \textbf{0.33}         \\ \hline
\end{tabular}
\label{tab:num_of_entries_method}
\end{table}

Firstly, we compare the size of the datasets expressed with the number of entries located in the datasets.
Table~\ref{tab:num_of_entries_method} shows the number of entries at method, class, and file level.
The number of entries contained in the traditional dataset are listed in the ``Trad'' column.
The ``BH'' column represents the number of entries in the BugHunter dataset, while ``Rate'' is calculated as follows:
$$Rate = \frac{\#\ of\ entries\ in\ the\ traditional\ dataset}{\#\ of\ entries\ in\ the\ BugHunter\ dataset}$$
The obtained rate is higher than 1.0 for most of the projects in case of the method level, which shows that the new approach contains less entries at this level.
A rate of \emph{1.54} is achieved at method level, \emph{0.41} at class level, and \emph{0.33} at file level.
It is important to note that the traditional dataset encompasses data for only a six-month long interval which has the most bugs in it.
On the other hand, the BugHunter dataset contains information from the beginning of the project up to September, 2017.
One would expect that the new approach will contain less entries than the traditional one since the BugHunter dataset contains only the entries which were affected by a closed bug.
However, the traditional dataset only depends on the size (number of files, classes, and methods) of the projects included.
In contrast, the BugHunter dataset highly depends on the number of closed bugs in the system (large projects can have small amount of reported bugs).
Even if no feature development was performed on a project (the size of the project remains quasi-same: in general no new files and classes are added, only modified) the number of closed bugs imply more entries in the BugHunter dataset, while size is not affected in any way in the traditional dataset.

To sum up, we cannot clearly decide whether the new dataset is more compact; however it is clearly visible that BugHunter could compress the bug related information at method level.
We achieved an F-measure value of 0.6319 at method level (see~Table~\ref{tab:pred_cap_method}) and the composed dataset contains 58,464 less entries than the traditional one.
In both datasets, the number of entries are sufficient to build a predictive model from; however we should investigate the predictive capabilities first to conclude our findings related to expressive power and compactness.

In the following, we present tables that capture the differences of the prediction capabilities between the two datasets (using F-measures, as before).
Tables~\ref{tab:pred_cap_method},~\ref{tab:pred_cap_class},~\ref{tab:pred_cap_file}, and~\ref{tab:pred_cap_projected} present machine learning results for method, class, file levels and also the F-measure values for the projected method level predictors, respectively.
The complete tables are not presented here due to lack of space, however average, standard deviation, min, and max values are calculated and included in the tables which provide a general picture for the comparison.
The \texttt{appendix.zip} file supplied as an online appendix (see Section~\ref{sec:dataset}) contains the complete tables with all F-measure values.
For the sake of clarity, we describe how we obtained the averages presented here in the paper in detail.
First, since the traditional dataset consists of multiple versions with bugs from six-months long intervals, for each project we selected the version from the traditional dataset that has the most number of bugs assigned to them.
After collecting the machine learning results of the selected versions, we calculated average F-measure values for each algorithm we used.
Then we ranked the algorithms based on these averages and we selected the one with the highest average value.
We used this average value for the traditional dataset in the comparison.
From the BugHunter dataset, we used the average F-measure value of the previously selected algorithm calculated on the results obtained after applying the Subtract filtering. 
We performed this process for method, class, file, and projected levels separately.

\begin{table}[!h]
\caption{Predictive capabilities - Method Level}
\fontsize{8pt}{10pt}\selectfont
\def\arraystretch{1}
\centering
\setlength{\tabcolsep}{3pt}
\begin{tabular}{|c|r|r|r|r|}
\hline
\textbf{Dataset}     & \textbf{Avg.} & \textbf{Std.dev.} & \textbf{Min}  & \textbf{Max}    \\ \hline
				BugHunter    & 0.6319        & 0.0836            & 0.3376        & 0.7573          \\ \hline
				Traditional  & 0.7348        & 0.0789            & 0.4019        & 0.8339          \\ \hline
\end{tabular}
\label{tab:pred_cap_method}
\end{table}

\begin{table}[!h]
\caption{Predictive capabilities - Class Level}
\fontsize{8pt}{10pt}\selectfont
\def\arraystretch{1}
\centering
\setlength{\tabcolsep}{3pt}
\begin{tabular}{|c|r|r|r|r|}
\hline
\textbf{Dataset}     & \textbf{Avg.} & \textbf{Std.dev.} & \textbf{Min}  & \textbf{Max}    \\ \hline
				BugHunter    & 0.5685        & 0.0704            & 0.3572        & 0.7400          \\ \hline
				Traditional  & 0.7710        & 0.0869            & 0.3446        & 0.8331          \\ \hline
\end{tabular}
\label{tab:pred_cap_class}
\end{table}

\begin{table}[!h]
\caption{Predictive capabilities - File Level}
\fontsize{8pt}{10pt}\selectfont
\def\arraystretch{1}
\centering
\setlength{\tabcolsep}{3pt}
\begin{tabular}{|c|r|r|r|r|}
\hline
\textbf{Dataset}     & \textbf{Avg.} & \textbf{Std.dev.} & \textbf{Min}  & \textbf{Max}    \\ \hline
				BugHunter    & 0.5147        & 0.0749            & 0.3328        & 0.7741          \\ \hline
				Traditional  & 0.6058        & 0.1076            & 0.2882        & 0.8247          \\ \hline
\end{tabular}
\label{tab:pred_cap_file}
\end{table}

\begin{table}[!h]
\caption{Predictive capabilities - Projected}
\fontsize{8pt}{10pt}\selectfont
\def\arraystretch{1}
\centering
\setlength{\tabcolsep}{3pt}
\begin{tabular}{|c|r|r|r|r|}
\hline
\textbf{Dataset}     & \textbf{Avg.} & \textbf{Std.dev.} & \textbf{Min}  & \textbf{Max}    \\ \hline
				BugHunter    & 0.7405        & 0.0914            & 0.3178        & 0.8386          \\ \hline
				Traditional  & 0.7831        & 0.0716            & 0.4399        & 0.8825          \\ \hline
\end{tabular}
\label{tab:pred_cap_projected}
\end{table}

On the traditional dataset, the machine learning algorithms performed better, achieving higher F-measure values in every case.
The two kinds of datasets differ fundamentally, because they are constructed with two different methods.
For the traditional dataset, we divided the history of the projects into six-month long intervals by selecting release versions from the GitHub repository.
We collected the reported bugs from these intervals and we assigned the buggy source code elements to these release versions based on the bugfixes.
Then we used the state of the source code elements from these selected versions to assemble the bug dataset.
This method was used in several previous studies~\cite{bugprediction,gyimothy2005empirical}.
It is necessary, because the bugs are reported usually after releasing a version, thus at the time of the release there are too few bugs to construct a bug dataset.
For the bug assignment, we used a heuristic method - similar to other studies~\cite{gyimothy2005empirical} - where we assigned each bug to the latest selected version before the bug was reported into the issue tracking system.
This method leads to some uncertainty in the dataset, because it could happen that the bug is not yet present in the assigned version.
Table~\ref{tab:uncertainty_trad} shows some characteristics of this uncertainty.

\begin{table}[!h]
\caption{Uncertainty in the traditional dataset}
\fontsize{10pt}{12pt}\selectfont
\def\arraystretch{1}
\centering
\setlength{\tabcolsep}{5pt}
\begin{tabular}{|l|c|c|c|}
\hline
\textbf{Project}     & \textbf{\begin{tabular}[c]{@{}c@{}}Average \\ days\end{tabular}} & \textbf{\begin{tabular}[c]{@{}c@{}}Average commits\\ before reported\end{tabular}} & \textbf{\begin{tabular}[c]{@{}c@{}}Average commits\\ before fixed\end{tabular}} \\ \hline
Android-U.-I.-L.     & 78.78    & 179.04  & 22.82  \\ 
antlr4               & 83.73    & 94.83   & 66.21  \\ 
BroadleafCommerce    & 96.40    & 524.88  & 116.74 \\ 
ceylon-ide-eclipse   & 136.05   & 442.00  & 20.22  \\ 
elasticsearch        & 93.85    & 1,004.60 & 382.79 \\ 
hazelcast            & 84.61    & 1,905.88 & 143.54 \\ 
junit                & 91.94    & 76.71   & 171.09 \\ 
MapDB                & 102.09   & 150.47  & 25.06  \\ 
mcMMO                & 108.71   & 289.83  & 41.72  \\ 
mct                  & 64.00    & 203.00  & 55.93  \\ 
neo4j                & 39.53    & 535.77  & 189.30 \\ 
netty                & 83.65    & 411.60  & 48.96  \\ 
orientdb             & 99.21    & 568.76  & 179.30 \\ 
oryx                 & 63.00    & 104.42  & 3.40   \\ 
titan                & 51.35    & 65.91   & 59.85  \\ \hline
\end{tabular}
\label{tab:uncertainty_trad}
\end{table}

The second column is the average number of days elapsed between the date of the release and the date of the bugs reported.
The maximum that could occur is 180 days, because we used intervals around 6 months long.
We can see that these averages are quite high, the overall average is 85 days.
The third column is the average number of commits contributed to the project between the release commit an the date of the bugs reported.
These values vary for each project, because it depends on the developers' activity.
For some projects (Elasticsearch, Hazelcast) it could mean thousands of modifications before the bug was reported.
The more commits are performed, the higher the probability that the source code element became buggy after the release.
The fourth column shows the average number of commits performed from the time when the bug was reported and when the fix was applied.
These numbers are much smaller, which also demonstrates that bugs are fixed relatively fast.
This fast corrective behavior causes before and after fix states to be less different for the BugHunter approach.
Consequently, less difference in metric values makes building a precise prediction model more difficult.

\begin{figure}[htb!]
\begin{center}
\includegraphics[width=0.9\textwidth]{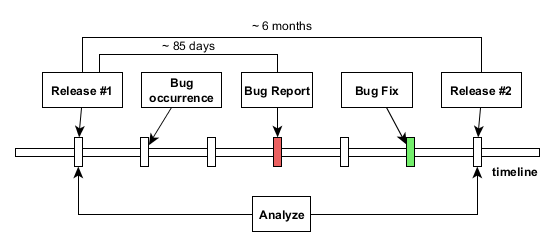}
\caption{Traditional approach} 
\label{fig:bh_vs_trad}
\end{center}
\end{figure}

\begin{figure}[htb!]
\begin{center}
\includegraphics[width=0.9\textwidth]{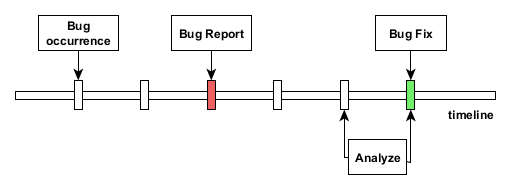}
\caption{BugHunter approach} 
\label{fig:bh_vs_trad2}
\end{center}
\end{figure}

The uncertainty is also visualized in Figures~\ref{fig:bh_vs_trad} and~\ref{fig:bh_vs_trad2} for the sake of better comprehension.
The first timeline (traditional dataset) shows a case when the actual bug occurred after Release~\#1, then the bug was reported and finally fixed before Release~\#2.
In this case, the source code elements touched by the bug are included in the dataset with a buggy label.
This dataset is created with the state captured at the time of Release~\#1.
However, these source code elements became buggy after Release~\#1, thus the dataset marked them buggy at that point incorrectly.
This error comes from the methodology itself which could be leveraged by narrowing the time window (which is 6 months wide traditionally).
On the other hand, 6 months interval is not an unwitting choice.
If we narrow down the time window, we will also have less bugs for an interval, which results in a more unbiased dataset, thus only less powerful predictive models can be built.
It would be important to see how many source code elements were marked as buggy incorrectly, however this cannot be easily measured since the exact time of the bug occurrence cannot be determined (we only know the time when a bug was reported).

The new BugHunter approach (see Figure~\ref{fig:bh_vs_trad2}), however, is free from the uncertainty mentioned above because it uses only the buggy and the fully fixed states of the bug related source code elements.
This way, the produced bug dataset is more precise, hence it is more appropriate for machine learning.
Therefore, we cannot clearly state that the traditional dataset is better, even despite the higher F-measure values.
The difference between the values of the two dataset is around 0.10 at method level, 0.21 at class level, and 0.09 at file level.
Projecting method level metrics to class level achieved almost as high an F-measure value (0.7405) as in the traditional case (0.7831).
The difference is only 0.04, yet it is on a much more precise dataset.

\vspace{6pt}
\setlength\parindent{0pt}
\fbox{
  \parbox{\textwidth - 15pt}{
		\textbf{Answering RQ3:} \textit{
		Traditional datasets include a high risk when labeling source code elements as buggy since the elements may become buggy after the release version.
		This injects false labeling into the training set, which might end up in deceptive machine learning results (as successfully predicting a bad label is not correct).
		Unfortunately, the number of incorrectly labeled source code elements cannot be determined since we only know the time when a bug was reported, we do not know the exact time when it was inserted into the system.
		These facts make it really hard to take one dataset and state that it is better for bug prediction.
		}
  }
}
\vspace{6pt}
\setlength\parindent{15pt}
	
\section{Threats to Validity}\label{threats}

In this section, we briefly describe the threats to validity.
Firstly, we present the construct validity, then the internal and external validity threats.

\subsection{Threats to Construct Validity}

When constructing a dataset in an automatic way, there are always some possible threats to validity.
We validated our matching algorithm on JUnit, which was fair in size.
However, investigating the validity of the matching in other systems could have revealed additional findings.

As we have seen, commit mismatches can occur during this process, which can distort the final bug dataset.
However, manually validating all bugs and the corresponding commits would have been an enormous task.

Deciding which source code elements are faulty and which are not can also cause a construct validity threat.
We consider a source code element faulty before the corresponding bug is fixed (the source code element had to be modified in that fix) and after the corresponding bug report is present.
The source code element can already be faulty before the report and can have multiple changes in that period; and it can be faulty also after the last fix, but we do not know the issue at that time, which can also distort the measurements.
Unfortunately, these uncertainties cannot be solved, since there are no further data to rely on.

\subsection{Threats to Internal Validity}

It would be meaningful to use multiple static source code analyzers in order to decrease the threats to internal validity caused by measuring source code element characteristics with only one tool.
However, it would mean much more work, and even then, additional manual validations would be needed to decide which tool measures a given metric more precisely, which often depends on interpreting the conceptual definitions.

\subsection{Threats to External Validity}

Currently, the constructed dataset consists of 15 projects which may limit the capabilities of the bug prediction models.
Selecting more projects to be included in the dataset would increase the generalizability of the built bug prediction models.
Considering additional source hosting platforms (SourceForge, Bitbucket, GitLab) would also increase the external validity of the dataset.

Widely used and accepted programming constructs and structures can vary from programming language to programming language.
Using different constructs may have a significant result on the calculated metric values.
Selecting projects written in different programming languages, not only Java software systems, could further strengthen the generalizability of our method.

\section{Conclusions and Future Work} \label{conclusion}

In this study, we developed a method that generates a bug dataset whose entries are source code elements touched by bugfixes mined from GitHub.
The entries represent before and after states of source code elements on which bug fixes were applied.
The presented approach allows the simultaneous processing of several publicly available projects located on GitHub, thereby resulting in the production of a large -- and automatically expandable -- dataset.
In contrast, previous studies have dealt with only a few large-scale datasets, which were created under strict individual management.
Additionally, our dataset contains new source code metrics compared to other datasets, allowing the examination of the relationship between these metrics and software bugs.
Furthermore, manual examinations showed the reliability of our approach, so the adaptation of project-specific labels to the presence of bugs remains the only non-automatic step.
Our initial adaptation of 15 suitable Java projects lead to the construction of the current dataset, which is one of our main -- publicly available -- contributions.

During empirical evaluations, we showed that the dataset can be used for further investigations such as bug prediction.
For this purpose, we used several machine learning algorithms at three different granularity levels (method, class, file) from which the method-level prediction achieved the highest F-measure values.
As a novel kind of experiment, we also investigated whether the method-level metrics projected to the class level are better predictors than the class-level metrics themselves, and found a significant improvement in the results.

As potential future work, we are planning to expand the dataset with additional projects and even additional data sources, such as SourceForge and Bitbucket.
Supporting different external bug tracking systems is another option for extending our approach.
We will also dedicate more attention to the concrete prediction models we generate, as this study focused solely on showing the conceptual usability of our dataset.

\section*{Acknowledgment}

This research was supported by the EU-supported Hungarian national grant GINOP-2.3.2-15-2016-00037 titled ``Internet of Living Things'' and by grant TUDFO/47138-1/2019-ITM of the Ministry for Innovation and Technology, Hungary.

\section*{Appendix}\label{sec:appendix}

\begingroup
  \setlength{\tabcolsep}{3pt} 
  \renewcommand{\arraystretch}{1}
\begin{sidewaystable}[!hp]
    \centering
    \caption{Significance test results for method level - Algorithms}
    \resizebox{\textwidth}{!}{
			\begin{tabular}{|l|r|r|r|r|r|r|r|r|r|r|}
    \hline
          & \multicolumn{1}{c|}{\textit{NaiveBayes}} & \multicolumn{1}{c|}{\textit{NaiveBayesMultinomial}} & \multicolumn{1}{c|}{\textit{Logistic}} & \multicolumn{1}{c|}{\textit{SGD}} & \multicolumn{1}{c|}{\textit{SimpleLogistic}} & \multicolumn{1}{c|}{\textit{VotedPerceptron}} & \multicolumn{1}{c|}{\textit{DecisionTable}} & \multicolumn{1}{c|}{\textit{OneR}} & \multicolumn{1}{c|}{\textit{J48}} & \multicolumn{1}{c|}{\textit{RandomForest}}\\
    \hline
    \textit{NaiveBayesMultinomial} & 0.900  (1.2814) & \gcell & \gcell & \gcell & \gcell & \gcell & \gcell & \gcell & \gcell & \gcell  \\
    \hline
    \textit{Logistic} & 0.129 (4.0704) & \textbf{0.007} (5.352)  & \gcell & \gcell & \gcell & \gcell & \gcell & \gcell & \gcell & \gcell \\
    \hline
    \textit{SGD} & \textbf{0.045} (4.5980) & \textbf{0.002} (5.8795) & 0.900 (0.5276) & \gcell & \gcell & \gcell & \gcell & \gcell & \gcell & \gcell \\
    \hline
    \textit{SimpleLogistic} & 0.053 (4.5227) & \textbf{0.002}(5.8041)  & 0.900 (0.4523) & 0.900 (0.0754) & \gcell & \gcell & \gcell & \gcell & \gcell & \gcell \\
    \hline
    \textit{VotedPerceptron} & 0.900 (1.8844) & 0.900 (0.6030) & \textbf{0.001} (5.9548) & \textbf{0.001}(6.4825)  & \textbf{0.001} (6.4071) & \gcell & \gcell & \gcell & \gcell & \gcell \\
    \hline
    \textit{DecisionTable} & \textbf{0.045} (4.5980) & \textbf{0.002} (5.8795) & 0.900 (0.5276) & 0.900 (0.0000) & 0.900 (0.0754) & \textbf{0.001} (6.4825) & \gcell & \gcell & \gcell & \gcell \\
    \hline
    \textit{OneR} & 0.302 (3.5428) & \textbf{0.027} (4.8242) & 0.900 (0.5276) & 0.900 (1.0553) & 0.900 (0.9799) & \textbf{0.006} (5.4272) & 0.900 (1.0553) & \gcell & \gcell & \gcell \\
    \hline
    \textit{J48} & \textbf{0.023} (4.8995) & \textbf{0.001} (6.1810) & 0.900 (0.8292) & 0.900 (0.3015) & 0.900 (0.3769) & \textbf{0.001} (6.7840) & 0.900 (0.3015) & 0.900 (1.3568) & \gcell & \gcell \\
    \hline
    \textit{RandomForest} & \textbf{0.001} (7.6131) & \textbf{0.001} (8.8943) & 0.302 (3.5426) & 0.546 (3.0151) & 0.513 (3.0905) & \textbf{0.001} (9.4976) & 0.546 (3.0151) & 0.129 (4.0704) & 0.679 (2.7136) & \gcell \\
    \hline
    \textit{RandomTree} & \textbf{0.019} (4.9749) & \textbf{0.001} (6.2564) & 0.900 (0.9045) & 0.900 (0.3769) & 0.900 (0.4523) & \textbf{0.001} (6.8594) & 0.900 (0.3769) & 0.900 (1.4322) & 0.900 (0.0754) & 0.712 (2.6382) \\
    \hline
    \end{tabular}
    }
    \label{tab:sign_res_method_ml}
  \end{sidewaystable}
\endgroup

  \begingroup
  \setlength{\tabcolsep}{3pt} 
  \renewcommand{\arraystretch}{1}
  \begin{sidewaystable}[htbp]
    \centering
    \caption{Significance test results for class level - Algorithms}
    \resizebox{\textwidth}{!}{
			\begin{tabular}{|l|r|r|r|r|r|r|r|r|r|r|}
    \hline
          & \multicolumn{1}{c|}{\textit{NaiveBayes}} & \multicolumn{1}{c|}{\textit{NaiveBayesMultinomial}} & \multicolumn{1}{c|}{\textit{Logistic}} & \multicolumn{1}{c|}{\textit{SGD}} & \multicolumn{1}{c|}{\textit{SimpleLogistic}} & \multicolumn{1}{c|}{\textit{VotedPerceptron}} & \multicolumn{1}{c|}{\textit{DecisionTable}} & \multicolumn{1}{c|}{\textit{OneR}} & \multicolumn{1}{c|}{\textit{J48}} & \multicolumn{1}{c|}{\textit{RandomForest}} \\
    \hline
    \textit{NaiveBayesMultinomial} & 0.900 (1.1307) & \gcell & \gcell & \gcell & \gcell & \gcell & \gcell & \gcell & \gcell & \gcell \\
    \hline
    \textit{Logistic} & 0.169 (3.9196) & 0.646 (2.7890) & \gcell & \gcell & \gcell & \gcell & \gcell & \gcell & \gcell & \gcell \\
    \hline
    \textit{SGD} & 0.169 (3.9196) & 0.646 (2.7890) & 0.900 (0.0000) & \gcell & \gcell & \gcell & \gcell & \gcell & \gcell & \gcell \\
    \hline
    \textit{SimpleLogistic} & \textbf{0.019} (4.9749) & 0.191 (3.8443) & 0.900 (1.0553) & 0.900 (1.0553) & \gcell & \gcell & \gcell & \gcell & \gcell & \gcell \\
    \hline
    \textit{VotedPerceptron} & 0.900 (1.8844) & 0.546 (3.0151) & \textbf{0.002} (5.8041) & \textbf{0.002} (5.8041) & \textbf{0.001} (6.8594) & \gcell & \gcell & \gcell & \gcell & \gcell \\
    \hline
    \textit{DecisionTable} & \textbf{0.004} (5.5780) & 0.063 (4.4473) & 0.900 (1.6583) & 0.900 (1.6583) & 0.900 (0.6030) & \textbf{0.001} (7.4624) & \gcell & \gcell & \gcell & \gcell \\
    \hline
    \textit{OneR} & 0.878 (2.2613) & 0.900 (1.1307) & 0.900 (1.6583) & 0.900 (1.6583) & 0.679 (2.7136) & 0.113 (4.1458) & 0.406 (3.3166) & \gcell & \gcell & \gcell \\
    \hline
    \textit{J48} & 0.369 (3.3920) & 0.878 (2.2613) & 0.900 (0.5276) & 0.900 (0.5276) & 0.900 (1.5829) & \textbf{0.009} (5.2764) & 0.900 (2.1860) & 0.900 (1.1307) & \gcell & \gcell \\
    \hline
    \textit{RandomForest} & 0.900 (0.7538) & 0.900 (0.3769) & 0.479 (3.1659) & 0.479 (3.1659) & 0.098 (4.2212) & 0.712 (2.6382) & \textbf{0.027} (4.8242) & 0.900 (1.5076) & 0.712 (2.6382) & \gcell \\
    \hline
    \textit{RandomTree} & 0.900 (1.6583) & 0.900 (0.5276) & 0.878 (2.2613) & 0.878 (2.2613) & 0.406 (3.3166) & 0.302 (3.5428) & 0.169 (3.9196) & 0.900 (0.6030) & 0.900 (1.7337) & 0.900 (0.9045) \\
    \hline
    \end{tabular}
      }
      \label{tab:sign_res_class_ml}
    \end{sidewaystable}
    \endgroup

    \begingroup
  \setlength{\tabcolsep}{3pt} 
  \renewcommand{\arraystretch}{1}
  \begin{sidewaystable}[htbp]
    \centering
    \caption{Significance test results for projected - Algorithms}
    \resizebox{\textwidth}{!}{
						\begin{tabular}{|l|r|r|r|r|r|r|r|r|r|r|}
    \hline
          & \multicolumn{1}{c|}{\textit{NaiveBayes}} & \multicolumn{1}{c|}{\textit{NaiveBayesMultinomial}} & \multicolumn{1}{c|}{\textit{Logistic}} & \multicolumn{1}{c|}{\textit{SGD}} & \multicolumn{1}{c|}{\textit{SimpleLogistic}} & \multicolumn{1}{c|}{\textit{VotedPerceptron}} & \multicolumn{1}{c|}{\textit{DecisionTable}} & \multicolumn{1}{c|}{\textit{OneR}} & \multicolumn{1}{c|}{\textit{J48}} & \multicolumn{1}{c|}{\textit{RandomForest}} \\
    \hline
    \textit{NaiveBayesMultinomial} & 0.900 (0.6784) & \gcell & \gcell & \gcell & \gcell & \gcell & \gcell & \gcell & \gcell & \gcell \\
    \hline
    \textit{Logistic} & \textbf{0.003} (5.7287) & \textbf{0.001} (6.4071) & \gcell & \gcell & \gcell & \gcell & \gcell & \gcell & \gcell & \gcell \\
    \hline
    \textit{SGD} & \textbf{0.001} (7.0101) & \textbf{0.001} (7.6885) & 0.900 (1.2814) & \gcell & \gcell & \gcell & \gcell & \gcell & \gcell & \gcell\\
    \hline
    \textit{SimpleLogistic} & \textbf{0.006} (5.4272) & \textbf{0.001} (6.1056) & 0.900 (0.3015) & 0.900  (1.5829) & \gcell & \gcell & \gcell & \gcell & \gcell & \gcell \\
    \hline
    \textit{VotedPerceptron} & 0.900 (1.8091) & 0.778 (2.4875) & 0.169 (3.9196) & \textbf{0.011} (5.2011) & 0.271 (3.6181) & \gcell & \gcell & \gcell & \gcell & \gcell \\
    \hline
    \textit{DecisionTable} & \textbf{0.001} (7.6885) & \textbf{0.001} (8.3669) & 0.900 (1.9598) & 0.900 (0.6784) & 0.878 (2.2613) & \textbf{0.002} (5.8795) & \gcell & \gcell & \gcell & \gcell \\
    \hline
    \textit{OneR} & \textbf{0.005} (5.5026) & \textbf{0.001} (6.1810) & 0.900 (0.2261) & 0.900 (1.5076) & 0.900 (0.0754) & 0.242 (3.6935) & 0.900 (2.1860) & \gcell & \gcell & \gcell \\
    \hline
    \textit{J48} & \textbf{0.001} (5.9548) & \textbf{0.001} (6.6332) & 0.900 (0.2261) & 0.900 (1.0553) & 0.900 (0.5276) & 0.113 (4.1458) & 0.900 (1.7337) & 0.900 (0.4523) & \gcell & \gcell \\
    \hline
    \textit{RandomForest} & \textbf{0.001} (7.4624) & \textbf{0.001} (8.1408) & 0.900 (1.7337) & 0.900 (0.4523) & 0.900 (2.0352) & \textbf{0.003} (5.6533) & 0.900 (0.2261) & 0.900 (1.9598) & 0.900 (1.5076) & \gcell \\
    \hline
    \textit{RandomTree} & \textbf{0.005} (5.5026) & \textbf{0.001} (6.1810) & 0.900 (0.2261) & 0.900 (1.5076) & 0.900 (0.0754) & 0.242 (3.6935) & 0.900 (2.1860) & 0.900 (0.0000) & 0.900 (0.4523) & 0.900 (1.9598) \\
    \hline
    \end{tabular}
      }
      \label{tab:sign_res_projected_ml}
    \end{sidewaystable}
    \endgroup

\begingroup
\setlength{\tabcolsep}{3pt} 
\renewcommand{\arraystretch}{1}
\begin{sidewaystable}[htbp]
  \centering
  \caption{Significance test results for file level - Algorithms}
  \resizebox{\textwidth}{!}{
				\begin{tabular}{|l|r|r|r|r|r|r|r|r|r|r|}
    \hline
          & \multicolumn{1}{c|}{\textit{NaiveBayes}} & \multicolumn{1}{c|}{\textit{NaiveBayesMultinomial}} & \multicolumn{1}{c|}{\textit{Logistic}} & \multicolumn{1}{c|}{\textit{SGD}} & \multicolumn{1}{c|}{\textit{SimpleLogistic}} & \multicolumn{1}{c|}{\textit{VotedPerceptron}} & \multicolumn{1}{c|}{\textit{DecisionTable}} & \multicolumn{1}{c|}{\textit{OneR}} & \multicolumn{1}{c|}{\textit{J48}} & \multicolumn{1}{c|}{\textit{RandomForest}} \\
    \hline
    \textit{NaiveBayesMultinomial} & 0.900 (1.8844) & \gcell & \gcell & \gcell & \gcell & \gcell & \gcell & \gcell & \gcell & \gcell \\
    \hline
    \textit{Logistic} & 0.129 (4.0704) & 0.900 (2.1860) & \gcell & \gcell & \gcell & \gcell & \gcell & \gcell & \gcell & \gcell \\
    \hline
    \textit{SGD} & 0.900 (0.1508) & 0.900 (2.0352) & 0.098 (4.2212) & \gcell & \gcell & \gcell & \gcell & \gcell & \gcell & \gcell \\
    \hline
    \textit{SimpleLogistic} & 0.148 (3.9950) & 0.900 (2.1106) & 0.900 (0.0754)  & 0.113 (4.1458)& \gcell & \gcell & \gcell & \gcell & \gcell & \gcell \\
    \hline
    \textit{VotedPerceptron} & 0.900 (1.2060) & 0.513 (3.0905) & \textbf{0.009} (5.2764) & 0.900 (1.0553) & \textbf{0.011} (5.2011) & \gcell & \gcell & \gcell & \gcell & \gcell \\
    \hline
    \textit{DecisionTable} & 0.443 (3.2412) & 0.900 (1.3568) & 0.900 (0.8292) & 0.369 (3.3920) & 0.900 (0.7538) & 0.063 (4.4473) & \gcell & \gcell & \gcell & \gcell \\
    \hline
    \textit{OneR} & 0.148 (3.9950) & 0.900 (2.1106) & 0.900 (0.0754) & 0.113 (4.1458) & 0.900 (0.0000) & \textbf{0.011} (5.2011) & 0.900 (0.7538) & \gcell & \gcell & \gcell \\
    \hline
    \textit{J48} & 0.513 (3.0905) & 0.900 (1.2060) & 0.900 (0.9799) & 0.443 (3.2412) & 0.900 (0.9045) & 0.085 (4.2965) & 0.900 (0.1508) & 0.900 (0.9045) & \gcell & \gcell \\
    \hline
    \textit{RandomForest} & 0.098 (4.2212) & 0.845 (2.3367) & 0.900 (0.1508) & 0.073 (4.3719) & 0.900 (0.2261) & \textbf{0.006} (5.4272) & 0.900 (0.9799) & 0.900 (0.2261) & 0.900 (1.1307) & \gcell \\
    \hline
    \textit{RandomTree} & 0.098 (4.2212) & 0.845 (2.3367) & 0.900 (0.1508) & 0.073 (4.3719) & 0.900 (0.2261) & \textbf{0.006} (5.4272) & 0.900 (0.9799) & 0.900 (0.2261) & 0.900 (1.1307) & 0.900 (0.0000) \\
    \hline
    \end{tabular}
    
    }
    \label{tab:sign_res_file_ml}
  \end{sidewaystable}
  \endgroup
  
\pagebreak

\bibliographystyle{elsarticle-num}
\bibliography{main}

\end{document}